\documentclass[
aps,pra,
reprint
]{revtex4-1}

\usepackage{amsmath}
\usepackage{amsfonts}
\usepackage{url}
\usepackage{bm}
\usepackage{bbold}
\usepackage{graphics}
\usepackage{braket}
\usepackage{paralist}

\begin{document}

\title{Density matrix quantum Monte Carlo}

\author{N. S. Blunt}
\affiliation{Department of Physics, Imperial College London, Exhibition Road, London, SW7 2AZ, U.K.}
\author{T. W. Rogers}
\affiliation{Department of Physics, Imperial College London, Exhibition Road, London, SW7 2AZ, U.K.}
\author{J. S. Spencer}
\affiliation{Department of Materials, Imperial College London, Exhibition Road, London, SW7 2AZ, U.K.}
\affiliation{Department of Physics, Imperial College London, Exhibition Road, London, SW7 2AZ, U.K.}
\author{W. M. C. Foulkes}
\affiliation{Department of Physics, Imperial College London, Exhibition Road, London, SW7 2AZ, U.K.}

\begin{abstract}
We present a quantum Monte Carlo method capable of sampling the full density matrix of a many-particle system at finite temperature.  This allows arbitrary reduced density matrix elements and expectation values of complicated non-local observables to be evaluated easily.  The method resembles full configuration interaction quantum Monte Carlo but works in the space of many-particle operators instead of the space of many-particle wave functions.  One simulation provides the density matrix at all temperatures simultaneously, from $T=\infty$ to $T=0$, allowing the temperature dependence of expectation values to be studied.  The direct sampling of the density matrix also allows the calculation of some previously inaccessible entanglement measures.
We explain the theory underlying the method, describe the algorithm, and introduce an importance-sampling procedure to improve the stochastic efficiency.  To demonstrate the potential of our approach, the energy and staggered magnetization of the isotropic antiferromagnetic Heisenberg model on small lattices, the concurrence of one-dimensional spin rings, and the Renyi $S_2$ entanglement entropy of various sublattices of the $6 \times 6$ Heisenberg model are calculated.  The nature of the sign problem in the method is also investigated.
\end{abstract}

\maketitle

\section{Introduction}
\label{sec:intro}
Quantum Monte Carlo (QMC) methods are well established as vital tools in the study of complex many-body quantum systems, often providing highly accurate results.  Projector methods such as diffusion Monte Carlo~\cite{Hammond1994,Foulkes2001} (DMC) and Green's function Monte Carlo~\cite{Kalos1962,Ceperley1979} (GFMC) grant access to zero-temperature properties by stochastically applying a projection operator to a starting wave function to obtain a statistical sampling of the ground state.  The fixed-node approximation~\cite{Anderson1975,Anderson1976,Moskowitz1982,Reynolds1982} allows projector QMC to be applied to systems with sign problems and often produces very good results, but its accuracy depends on the quality of the trial nodal surface and is difficult to assess.  Another drawback of projector QMC is that expectation values of quantum mechanical observables that do not commute with the Hamiltonian are difficult to calculate~\cite{Hammond1994,Foulkes2001,Gaudoin2007,Casulleras95}.  

Finite-temperature QMC methods take a different approach.  Path-integral Monte Carlo (PIMC) calculations express the partition function, $Z=\textnormal{Tr}(e^{-\beta H})$, as a sum of contributions from paths through Hilbert space~\cite{Feynman1953}.  With an appropriate update procedure, the paths can be sampled with the correct probabilities, thus allowing finite-temperature expectation values to be evaluated.  The stochastic series expansion (SSE) method~\cite{Sandvik2010} has much in common with PIMC\@.  These algorithms also allow access to ground-state properties in principle, but in practice the sign problem is often insurmountable at low temperatures.

The full configuration interaction quantum Monte Carlo (FCIQMC) method recently introduced by Booth, Thom and Alavi~\cite{Booth2009, Booth2013} is a projector method for studying zero-temperature properties, and, as such, has much in common with DMC and GFMC\@. However, unlike DMC and GFMC, where the sampling of the ground-state wave function is performed in real space, FCIQMC samples the components of the wave function in a discrete basis.  Crucially, no prior knowledge of the nodal structure of the ground-state wave function is required to reach the exact ground state. Rather, the sign problem manifests itself in the large but system-specific population of quantum Monte Carlo walkers required in order for the ground state of the Hamiltonian to emerge~\cite{Spencer2012} from the background noise. The system sizes accessible to FCIQMC are limited by the amount of memory available to store these walkers.  However, the method has proven highly successful in many chemical systems, reducing the memory needed to achieve FCI-quality results by several orders of magnitude~\cite{Booth2010,Cleland2010,Cleland2012,Shepherd2012PRB,Daday2012}. This has led to much interest in this direction and research into fundamental improvements and new applications of the algorithm continues~\cite{Cleland2010,Petruzielo2012}.

This article presents a closely-related QMC method, which we call density matrix quantum Monte Carlo (DMQMC). Like the path-integral and SSE methods, DMQMC allows finite-temperature results to be calculated.  However, it uses a projection approach to achieve this and thus has more in common with zero-temperature QMC methods. DMQMC was inspired by FCIQMC and shares many of its features, but samples the elements of the density matrix instead of the components of the wave function in a discrete basis. This enables expectation values of arbitrary quantum mechanical observables to be calculated easily, even when the operator corresponding to the observable does not commute with the Hamiltonian.  Such expectation values are difficult to calculate using other QMC methods~\cite{Hammond1994,Foulkes2001,Gaudoin2007,Casulleras95}.  Moreover, the ability to directly sample the density matrix means that many quantum information measures are accessible.  These advantages cannot be expected to come without drawbacks, and, indeed, the systems to which we have successfully applied DMQMC to date are small by the standards of other finite-temperature methods. However, the potential uses of a direct stochastic sampling of the density matrix are such that DMQMC deserves investigation.  This paper demonstrates the use of DMQMC by studying the antiferromagnetic Heisenberg model, but DMQMC is a general method, applicable to bosonic and fermionic systems and to real molecules and solids.

Section~\ref{sec:FCIQMC} summarizes the FCIQMC method, setting the stage for the description of DMQMC in Section~\ref{sec:DMQMC}. In Section~\ref{sec:importance_sampling} an importance-sampling procedure is introduced.  The DMQMC method is then applied to the isotropic antiferromagnetic Heisenberg model in Section~\ref{sec:results} and used to calculate the energy and staggered magnetization of small square lattices, the concurrence of one-dimensional rings, and the Renyi $S_2$ entanglement entropy of various sublattices of the $6 \times 6$ lattice. The sign problem in DMQMC is investigated in Section~\ref{sec:sign_problem}.  We discuss our results and offer some concluding remarks in Section~\ref{sec:discussion}.  Hartree atomic units are used throughout.

\section{Full Configuration Interaction Quantum Monte Carlo}
\label{sec:FCIQMC}
The DMQMC algorithm was formulated in analogy with the FCIQMC method. The two methods share many of the same features and it is often useful to compare them.  We therefore begin by providing a brief summary of the FCIQMC method. For more detailed discussions readers are referred to Refs.~(\onlinecite{Booth2009}) and (\onlinecite{Spencer2012}).

Consider the imaginary-time Schr\"odinger equation
\begin{equation}
\frac{d \ket{\Psi}}{d \tau} = -\hat{H}\ket{\Psi}.
\end{equation}
The general solution to this equation is
\begin{equation}
\ket{\Psi(\tau)} = e^{-\tau \hat{H}} \ket{\Psi(\tau=0)},
\end{equation}
for some initial wave function $\ket{\psi(\tau=0)}$. If the initial wave function has a non-zero ground-state component, $c_0(0)$, it is easy to see that $\ket{\Psi(\tau)}$ will become proportional to the ground state in the limit of large $\tau$,
\begin{equation}
\ket{\Psi(\tau \rightarrow \infty)} = c_0(0) e^{-\tau E_0} \ket{E_0},
\label{eq:groundstateProjection}
\end{equation}
where $\ket{E_0}$ is the ground state and $E_0$ the ground-state energy. The factor of $e^{-E_0 \tau}$ can be removed by choosing the zero of energy such that $E_0 = 0$. In practice, since $E_0$ is usually unknown, we solve
\begin{equation}
\label{eq:FCIQMCEquation}
\frac{d \ket{\Psi}}{d \tau} = -(\hat{H} - S\hat{\mathbb{1}})\ket{\Psi} = \hat{T} \ket{\Psi},
\end{equation}
where we have defined $\hat{T} = -(\hat{H}-S\hat{\mathbb{1}})$. The energy shift $S$ is adjusted slowly during the simulation to keep the normalization approximately constant.  The long-time average of $S$ provides a measure of the ground-state energy.

The above theory is common to all projector methods; the difference is how they achieve the evolution to the ground state. FCIQMC works in a discrete basis of kets $\ket{X_i}$, which are normally Slater determinants for fermions or permanents for bosons. The components of the wave function in this basis are represented stochastically by a collection of markers. Following Anderson~\cite{Anderson1975,Anderson1976}, we refer to these markers as ``psips''. Each psip has an associated sign, $q = \pm 1$, which we refer to as its ``charge'', and resides on a particular basis state $\ket{X_i}$ (or on ``site $i$''). The expected value of the net psip charge on a basis state is proportional to the amplitude of that state in the expansion of the wave function. At any point in a simulation, the distribution of psip charges does not need to provide an accurate representation of the wave function. Rather, the FCIQMC method only requires that the \emph{expectation values} of the site charges represent the ground state~\cite{Cleland2012}. Thus, at any point in the simulation, the memory required to sample the wave function may be many orders of magnitude smaller than the memory required to store the whole state.

Booth and co-workers~\cite{Booth2009} introduced an algorithm to evolve the population of psips according to the imaginary-time Schr\"odinger equation. This can be summarized as follows.  For each time step $\Delta\tau$ we loop over the entire population of psips and perform the following steps:
\begin{enumerate}
\item \emph{Spawning:} Allow a psip with charge $q_i$ on site $i$ to spawn onto connected sites $j$, where $T_{ij} \neq 0$ and $i \not= j$, with probability $\lvert T_{ji} \rvert \Delta\tau$. If the spawning attempt is successful, a psip is born at site $j$ with charge $q_j = \textnormal{sign}(T_{ji})q_i$.
\item \emph{Diagonal death/cloning:} Each psip has the chance to either clone or die with probability $\lvert T_{ii} \rvert \Delta\tau$.  The consequence of a successful death/cloning event depends on the sign of the diagonal matrix element: if $T_{ii} > 0$ the psip is cloned; otherwise the psip is removed from the simulation.
\item \emph{Annihilation:} Pairs of psips on the same site with opposite charges cancel out (``annihilate'') and are removed from the simulation, leaving a population of only a single charge type on each site.
\end{enumerate}
The FCIQMC algorithm samples the solution of a first-order Euler finite-difference approximation to Eq.~(\ref{eq:FCIQMCEquation}). Hence, the distribution of psips gives a stochastic representation of the wave function and, as $\tau \to \infty$, the psips settle down to sample the ground-state wave function.

The psip annihilation step does not alter the total charge on a site. However, it has been shown to be vital in order for the true ground-state wave function to emerge in systems with sign problems~\cite{Spencer2012}. Similar and more complex walker cancellation mechanisms have been attempted in continuum DMC and GFMC calculations with less success~\cite{PhysRevLett.85.3547, PhysRevC.32.1735, anderson:7418, PhysRevLett.67.3074, PhysRevE.50.3220, PhysRevE.53.5420}. Walker cancellation in FCIQMC works better because psips are more likely to encounter each other in a discrete and finite Hilbert space.

The ability of FCIQMC to tackle the sign problem through the annihilation step is perhaps its most significant advantage. Annihilation leads to the characteristic population dynamics and allows the true ground state to emerge without any knowledge of the nodal structure of the ground-state wave function~\cite{Spencer2012}.  At the start of an FCIQMC simulation, the shift is held constant at a value large enough to ensure that the psip population grows exponentially.  Eventually, when a system-specific population of psips is reached, the annihilation rate becomes equal to the spawning rate and the population spontaneously plateaus. During this plateau period the ground state of the Hamiltonian emerges, after which the population begins to grow again.  The population can then be controlled by varying the shift.  The psip population at the plateau must be a small fraction of the number of basis states in the Hilbert space in order for FCIQMC to be more (memory) efficient than an exact diagonalization.

The ground-state energy in FCIQMC can be calculated using a ``projected estimator'' based on the expression
\begin{align}
E_0 = \frac{\bra{\Psi^T} \hat{H} \ket{E_0}}{\langle\Psi^T|E_0\rangle} = \frac{\sum_{i,j} \psi_i^T H_{ij} \chi_j}{\sum_i \psi_i^T \chi_i}.
\end{align}
Here $\ket{\Psi^T}$ is an appropriate trial function with components $\psi_i^T$ in the many-particle basis, $\ket{X_i}$, and $\chi_i$ is a component of the exact ground state in this basis.  FCIQMC was initially performed using only the Hartree--Fock determinant as the trial function; multi-reference trial functions have subsequently been shown to substantially decrease statistical fluctuations~\cite{Petruzielo2012}.  The ground-state energy is obtained by averaging the numerator and denominator separately, with the total psip charge on each site, $q_i^{\text{tot}}$, used in the place of corresponding exact amplitude, $\chi_i$.

Calculating the ground-state expectation value of an operator $\hat{O}$ that does not commute with the Hamiltonian is more difficult because a projected estimator cannot be used. Instead, assuming that $\ket{E_0}$ is real, it is necessary to evaluate
\begin{align}
\langle \hat{O} \rangle = \frac{\bra{E_0} \hat{O} \ket{E_0}}{\langle E_0|E_0\rangle} = \frac{\sum_{i,j} O_{ij} \chi_i \chi_j}{\sum_i \chi_i^2},
\end{align}
where $O_{ij}=\bra{X_i}\hat{O}\ket{X_j}$.  Although $\langle q_i^{\text{tot}} \rangle = \chi_i$, the expectation value of a product is not equal to the product of the expectation values: $\langle q_i^{\text{tot}} q_j^{\text{tot}} \rangle \not= \chi_i \chi_j$. This means that $\chi_i \chi_j$ \emph{cannot} be obtained by averaging the products of the instantaneous psip weights. One could in principle average $q_i^{\text{tot}}$ over many iterations to obtain $\chi_i$ before multiplying $\chi_i$ and $\chi_j$, but this would involve storing a number for every basis function, which is impractical due to memory limitations.

This problem is not easy to overcome and there is currently no way to evaluate general expectation values exactly and efficiently within the FCIQMC framework. Indeed, the calculation of general expectation values in other Monte Carlo methods is often a difficult task.

\section{Density Matrix Quantum Monte Carlo}
\label{sec:DMQMC}
We now show how an FCIQMC-like dynamics can be used to sample both finite-temperature and ground-state density matrices. We first consider the thermal density matrix and how it can be evolved as a function of inverse temperature by solving the symmetrized Bloch equation. We then draw upon analogies with FCIQMC to formulate the DMQMC algorithm before discussing the calculation of estimators for a general quantum mechanical observable. This section ends with an explanation of how to sample a reduced density matrix in order to calculate estimators of entanglement measures.

\subsection{Theory}
Since the psip population (and hence the normalization) varies during a quantum Monte Carlo simulation, it is convenient to work with the unnormalized thermal density matrix
\begin{equation}
\label{eq:Rho}
\hat{\rho}(\beta) = e^{-\beta \hat{H}},
\end{equation}
where $\hat{H}$ is the Hamiltonian operator and $\beta = 1/k_BT$ is the inverse temperature.  The canonical partition function $Z(\beta)$ is given by:
\begin{equation}
\label{eq:partitionFunction}
Z(\beta) = \textnormal{Tr}\left(\hat{\rho}(\beta)\right).
\end{equation}
Differentiating $\hat{\rho}(\beta)$ with respect to $\beta$ shows that it obeys both the Bloch equation,
\begin{equation}
\label{eq:unsymmetrisedBloch}
\frac{d \hat{\rho}}{d \beta} = -\hat{H}\hat{\rho},
\end{equation}
and the symmetrized Bloch equation~{\footnote{%
There is a close relationship between the symmetrized Bloch equation in Eq.~(\ref{eq:symmetrisedBloch}) and the von Neumann and quantum Liouville equations, $i\hbar \frac{\partial\rho}{\partial t} = [H,\rho]$.
The von Neumann equation may be derived by using the time-dependent Schr{\" o}dinger equation to work out how the arbitrary initial density matrix of Eq.~(\ref{eq:rho_mixed}) evolves.
Analogously, the Bloch equation may be derived by using the \emph{imaginary-time} Schr{\" o}dinger equation to work out how Eq.~(\ref{eq:rho_mixed}) evolves in imaginary time.
The switch from real to imaginary time replaces the commutator of $\hat{H}$ and $\rho$ appearing in the von Neumann equation with the anticommutator appearing in the Bloch equation.
The additional factor of $\frac{1}{2}$ is introduced for convenience: it ensures that evolving Eq.~(\ref{eq:symmetrisedBloch}) for $\beta$ units of imaginary time, starting from the unit operator, $\rho(\beta=0)=\hat{I}$, yields the canonical density matrix at inverse temperature $\beta$, not $2\beta$.
}},
\begin{equation}
\label{eq:symmetrisedBloch}
\frac{d \hat{\rho}}{d \beta} = -\frac{1}{2}(\hat{H}\hat{\rho} + \hat{\rho}\hat{H}) = -\frac{1}{2}\{\hat{H},\hat{\rho}\}.
\end{equation}

The symmetrized version turns out to be more useful for our purposes. Consider the general solution to Eq.~(\ref{eq:symmetrisedBloch}),
\begin{equation}
\label{eq:generalSolution}
\hat{\rho}(\beta) = e^{-\frac{\beta}{2} \hat{H}} \hat{\rho}(\beta=0) e^{-\frac{\beta}{2} \hat{H}},
\end{equation}
and let $\ket{E_i}$ denote an energy eigenstate with energy $E_i$. Without loss of generality, we assume that $E_0=0$. By expanding $\hat{\rho}(\beta=0)$ as
\begin{equation}
    \label{eq:rho_mixed}
\hat{\rho}(\beta=0) = \sum_{i,j} \rho_{ij}(0) \ket{E_i}\bra{E_j}
\end{equation}
and applying Eq.~(\ref{eq:generalSolution}) it is seen that $\hat{\rho}(\beta)$ becomes proportional to the ground-state density matrix as $\beta \to \infty$:
\begin{equation}
\hat{\rho}(\beta \to \infty) = \rho_{00}(0) \ket{E_0}\bra{E_0}.
\end{equation}
A similar analysis shows that the ground-state density matrix is \emph{not} reached if a general initial density matrix is propagated using the unsymmetrized Bloch equation.  Hence, in cases where ground-state properties are desired, Eq.~(\ref{eq:symmetrisedBloch}) is more useful.  If we wish to sample thermal properties at any finite temperature $T$, we must also impose the correct initial condition on Eq.~(\ref{eq:symmetrisedBloch}). The most convenient is at $\beta = 0$, where 
\begin{equation}
\label{eq:initialCondition}
\hat{\rho}(\beta = 0) = \hat{\mathbb{1}}.
\end{equation}

Eqs.~(\ref{eq:symmetrisedBloch}) and (\ref{eq:initialCondition}) provide everything required to stochastically sample the thermal density matrix. As in FCIQMC we introduce a collection of psips, which in DMQMC occupy not basis states $\ket{X_i}$ but basis operators $\ket{X_i}\bra{X_j}$ of the expansion of the full many-particle density operator $\hat{\rho}(\beta)$ in some convenient basis set.  At the start of a simulation the psips are distributed randomly along the diagonal of the infinite-temperature density matrix $\rho_{ij}(\beta=0) = \delta_{ij}$. The psips then evolve under a set of rules so that they sample the matrix elements ${\rho}_{ij}(\beta)$ in the chosen basis at each iteration. In order to prevent the population of psips from exploding, we introduce an energy shift, $S$, and let $\hat{H} \to \hat{H} -S\hat{\mathbb{1}}$ as in FCIQMC\@.  Eq.~(\ref{eq:symmetrisedBloch}) thus becomes
\begin{equation}
\label{eq:symmetrisedBlochWithUpdateMatix}
\frac{d \hat{\rho}}{d \beta} = \frac{1}{2}(\hat{T}\hat{\rho} + \hat{\rho}\hat{T}),
\end{equation}
where $\hat{T} = -(\hat{H} - S\hat{\mathbb{1}})$ is the ``update matrix''.

\subsection{Algorithm}

We use the following algorithm to evolve a collection of psips according to Eq.~(\ref{eq:symmetrisedBlochWithUpdateMatix}).
For a single step in inverse temperature of $\Delta\beta$, we loop over the entire population of psips and perform the following steps:
\begin{enumerate}
\item \emph{Spawning along columns of the density matrix:} Allow a psip with charge $q_{ij}$ on site $(i,j)$ to spawn onto connected sites $(k,j)$, where $T_{ik} \ne 0$ and $i \ne k$, with probability $\frac{1}{2}\lvert T_{ik} \rvert \Delta\beta$. If the spawning attempt is successful, a psip is born at $(k,j)$ with charge $q_{kj} = \textnormal{sign}(T_{ik})q_{ij}$. 
\item \emph{Spawning along rows of the density matrix:} Similarly, allow a psip with charge $q_{ij}$ on site $(i,j)$ to spawn onto connected sites $(i,k)$ with probability $\frac{1}{2}\lvert T_{jk} \rvert \Delta\beta$. If the spawning attempt is successful, a psip is born at site $(i,k)$ with charge $q_{ik} = \textnormal{sign}(T_{jk})q_{ij}$.
\item \emph{Diagonal death/cloning:} If there is a psip on site $(i,j)$ and $T_{ii} + T_{jj} < 0$, then with probability $p_d = \frac{1}{2}|T_{ii}+T_{jj}|\Delta\beta$ that psip is killed and removed from the simulation; if $T_{ii} + T_{jj} > 0$, the psip is cloned (i.e., a new psip is created on the same site and with the same charge) with probability $p_d$.
\item \emph{Annihilation:} Pairs of psips inhabiting the same site $(i,j)$ with opposite charges annihilate and are removed from the simulation, leaving a population of only a single charge type on each site.
\end{enumerate}
The first three steps describe a stochastic algorithm to sample the solution of a first-order Euler finite-difference approximation to Eq.~(\ref{eq:symmetrisedBloch}). The distribution of psip charges at $\beta+\Delta\beta$ is thus proportional to the density matrix at this inverse temperature, provided that the distribution of charges at $\beta$ was correct.

The DMQMC method shares many similarities with FCIQMC, namely:
\begin{itemize}
    \item Annihilation does not alter the expected (normalized) psip distribution but serves to overcome the sign problem~\cite{Booth2009,Spencer2012}.
    \item The underlying finite-difference approximation is stable if $ 0 < \Delta\beta < 2/(E_{\text{max}} - E_0)$, where $E_{\text{max}}$ is the largest eigenvalue of the Hamiltonian matrix~\cite{Spencer2012}.  This is a sufficient condition to ensure correct projection onto the exact ground state, but the finite value of $\Delta\beta$ leads to an error of $\mathcal{O}(\Delta\beta)$ in the density matrix at temperatures greater than zero.  It is therefore necessary to check that finite-temperature results have converged with respect to $\Delta\beta$.
    \item The familiar shift-update algorithm already used in DMC~\cite{Umrigar1993} and FCIQMC~\cite{Booth2009} simulations is employed to modify $S$ and thus to control the population.  The shift, $S$, is adjusted according to
\begin{equation}
    \begin{split}
        S(\beta +& A\Delta\beta) =  S(\beta) \\ 
                                 &- \frac{\zeta}{A\Delta\beta}\ln\left( \frac{N_p(\beta + A\Delta\beta)}{N_p(\beta)}\right),
    \end{split}
\end{equation}
where $A$ is the number of $\beta$-steps between shift updates, $\zeta$ is a shift damping parameter, and $N_p(\beta)$ is the total number of psips at the inverse-temperature $\beta$. During simulations, $\zeta$ is chosen carefully to prevent large fluctuations in $S$.
    \item Rather than attempting all possible spawning events from a psip on a given site, it is computationally efficient to attempt just one (or a small number) of the many possible spawning events and reweight the acceptance probabilities accordingly~\cite{Booth2009}.
    \item The algorithm is highly parallelizable --- only the annihilation step requires communication between CPU cores and so using a large psip population is a viable option.
\end{itemize}

As with all projection methods, the ground state is approached as $\beta\to\infty$. Once convergence to the ground state has been attained the density matrix no longer depends on $\beta$ (remember that $S$ is chosen to keep the normalization fixed) and the statistical errors of measured ground-state expectation values can easily be reduced by averaging over many iterations. A single simulation is therefore sufficient to obtain accurate ground-state expectation values. The estimation of finite-temperature properties is more difficult because the inverse temperature $\beta$ changes continuously as the simulation progresses; it is not possible to hold the simulation at a specific temperature whilst statistics are accumulated. 

With access to a stochastic sampling of the unnormalized density matrix at a given temperature, the expectation value of any quantum mechanical observable, $\hat{O}$, at that temperature can be calculated using
\begin{equation}
    \langle \hat{O}\rangle = \frac{\textnormal{Tr}(\hat{\rho} \: \hat{O})}{\textnormal{Tr}(\hat{\rho})} = \frac{\sum_{i,j} \overline{q}^{\text{tot}}_{ij}O_{ji}}{\sum_i \overline{q}^{\text{tot}}_{ii}}.
\label{eq:estimator}
\end{equation} 
The numerator and denominator must be sampled and averaged separately at each temperature to achieve the desired statistical accuracy.  Hence a simulation involves repeatedly evolving the density matrix from $\beta=0$ to some chosen maximum value of $\beta$, a process we call a ``$\beta$-loop''.  This also gives us the freedom to allow Eq.~(\ref{eq:initialCondition}) to be satisfied only \emph{on average}, which greatly reduces the memory demands.  Each $\beta$-loop is initialized with a different random number seed and with psips randomly distributed with uniform probability along the diagonal of the density matrix.  Statistics are accumulated over a sufficiently large number of $\beta$-loops in order to obtain the desired statistical accuracy.  As each $\beta$-loop is completely independent, performing multiple $\beta$-loops is an embarrassingly parallel task and each $\beta$-loop gives statistically independent data points.

A primary concern is that averages over an impractically large number of $\beta$-loops may be required. Indeed, it is not uncommon to have to average over $\mathcal{O}(10^6)$ iterations in ground-state calculations.  In practice, however, reaching a sufficiently large number of $\beta$-loops does not appear to be a problem. The crucial difference from the ground-state case is that each $\beta$-loop provides statistically independent data whereas ground-state calculations need to overcome inherent correlations between nearby iterations.  Furthermore, as will be shown, the quality of estimates obtained using a given number of $\beta$-loops is better at high temperature than at low temperature.  Another advantage is that a single simulation can provide data across the entire temperature range being studied.

The energy shift, $S$, is varied in order to control the normalization.  This, however, introduces a bias: if the shift is varied by $\Delta S$, then the psip population is effectively altered by a factor $e^{\Delta \beta \Delta S}$ over the next $\beta$ step. As a result, configurations in which the average energy of the psip population happens to be less negative than usual are effectively given too large a weight.  This problem is particularly severe in DMQMC because the results at each temperature are obtained by averaging over separate $\beta$-loops, and thus, for each quantity contributing to a given estimator, the corresponding shift (and hence population) profiles can be very different.  

The population bias can be greatly reduced using a method suggested by Umrigar \emph{et al.}\ in the context of DMC~\cite{Umrigar1993}, in which each sampled quantity proportional to the psip population is multiplied by the factor
\begin{equation}
\Pi(\beta,B) = \prod^{\tilde{B}-1}_{m=0} e^{-S(\beta-m\Delta\beta)},
\label{bias_weights}
\end{equation}
where $B$ is some chosen number of factors and $\tilde{B}=\textnormal{min}(\frac{\beta}{\Delta \beta},B)$. By multiplying by this factor, we remove the last $\tilde{B}$ factors of $e^{\Delta \beta \Delta S}$ introduced by varying the shift. As $B \to \infty$ the population control bias should be completely removed. The population control bias can also be reduced by using a larger population of psips.

Another concern is that the number of elements in the density matrix is the square of the dimension of the Hilbert space of many-particle states.
At first glance it might seem that, without dramatically increasing the number of psips, the density matrix would be very poorly sampled. However, the dimension of the Hilbert space, $D$, rises exponentially with the number of particles or sites, $N$, so that $D \propto e^{\alpha N}$ and thus $D^2 \propto e^{\alpha(2N)}$. The doubling of the exponent implies that a DMQMC simulation for an $N/2$-site lattice model requires approximately the same number of psips as an $N$-site FCIQMC simulation to achieve the same sampling quality. Moreover, DMQMC estimators for operators that do not commute with the Hamiltonian often have significantly smaller variance than the forward-walking or other estimators required to evaluate the ground-state expectation values of such operators in FCIQMC.

\subsection{Entanglement measures}
\label{subsec:entanglementmeasures}

Entanglement measures are well established as an important concept in quantum information theory and have recently become a subject of active research in the condensed matter community. For example, changes in entanglement are observed at quantum phase transitions~\cite{Amico2008} and used to classify properties of Fermi liquids~\cite{McMinis2013,Swingle2013} and bonding in small molecules~\cite{Tubman2012}.  The Lanczos method can be used to calculate entanglement measures in small systems and the density matrix renormalization group method is applicable in one-dimensional systems~\cite{Hastings2010}, but the study of entanglement in large systems in more than one dimension is less straightforward. Relationships between reduced density matrices (RDMs) and spin correlation functions~\cite{Glaser2003} allow QMC methods to access some entanglement measures in certain situations~\cite{Roscilde2005}, and Sandvik recently introduced a QMC method formulated in the valence-bond basis~\cite{Sandvik2005} that allowed certain new entanglement entropy measures to be evaluated~\cite{Alet2007, Lin2010}. Hastings \emph{et al.}\ showed that Renyi $S_2$ entanglement entropy can also be calculated~\cite{Hastings2010}.    However, in general, the inability of QMC methods to directly access the density matrix and reduced density matrices of systems has hindered their use in this area.

Within DMQMC it straightforward to obtain a stochastic representation of any reduced density matrix element from a stochastic representation of the full density matrix.  
A RDM can be sampled by ``tracing out'' unwanted psips. Consider a composite quantum system $C$, which can be partitioned into two subsystems $A$ and $B$ so that
$\mathcal{H}_C = \mathcal{H}_A \otimes \mathcal{H}_B$, where $\mathcal{H}_A$ ($\mathcal{H}_B$) denotes the Hilbert space of system $A$ ($B$). The RDM $\rho_A$ that describes sublattice $A$ is defined by taking the partial trace of the full density matrix, $\rho_{C}$, over all the sites on sublattice $B$:
\begin{equation}
\rho_A = \textnormal{Tr}_B(\rho_{C}).
\end{equation}

Our implementation of DMQMC represents the many-particle basis functions as bit strings~\cite{Knowles1984}, where each bit refers to the state of a single spin.  To evaluate $\rho_A$ we construct a mask, $I_{B}$, which only has bits set which correspond to spins in subsystem $B$.  The $(i,j)$ density matrix element of $\rho_{C}$ contributes to $\rho_A$ if the result of the logical \texttt{AND} operation of the $i$ string with the $I_B$ mask is identical to that of the $j$ string.  The corresponding element in the RDM can be found by taking \texttt{AND} with an analogous $I_A$ mask.

    Evaluating the von Neumann entanglement entropy, $S_1 = -\textrm{Tr}(\rho \; \textrm{log} \; \rho)$, and other non-linear functions of a RDM $\rho$ is challenging because $\langle f(\rho) \rangle \not= f(\langle \rho \rangle )$. Thus, although it is easy to accumulate contributions to $\langle f(\rho)\rangle$ ``on the fly'', this does not yield the correct result $f(\langle \rho\rangle)$. A well-averaged estimate of the full RDM is required before the non-linear function can be calculated accurately. Accumulating this estimate requires storing the full RDM, limiting studies to relatively small subsystems. It would be reasonable to store the reduced density matrix of a subsystem of 15 spins, but much larger subsystems are out of reach. The von Neumann entanglement entropy is particularly challenging because its derivative with respect to any RDM eigenvalue tends to infinity as the eigenvalue tends to zero. Thus, small errors in small RDM eigenvalues can lead to large errors in the final estimate of $S_1$.

Two-qubit entanglement measures are more straightforward to calculate as the small size of the RDM allows it to be estimated accurately with relative ease. In this paper a concurrence estimate was taken from each $\beta$-loop by first averaging the relevant RDM in the ground state. These concurrence estimates were themselves averaged to provide a final value and associated error.

The replica trick approach, also used by Hastings \emph{et al.}~\cite{Hastings2010}, allows Renyi entropies to be calculated in a rigorously unbiased manner without storing and averaging the full RDM\@. Two DMQMC calculations performed simultaneously, each starting from a different random number seed and thus following a different Markov chain, are statistically independent. The Renyi-2 entropy can therefore be evaluated in an unbiased fashion:
\begin{equation}
S_2 = -\textrm{log}_2( \langle \sum_{i,j} q_{ij}^{tot} w_{ij}^{tot} \rangle ),
\label{r2_dmqmc}
\end{equation}
where $q_{ij}^{tot}$ ($w_{ij}^{tot}$) is the total psip amplitude on the $ij$-th element of the reduced density matrix in the first (second) simulation.
Thus, by averaging $\sum_{ij} q_{ij}^{tot} w_{ij}^{tot}$ at each temperature over many $\beta$-loops, it is straightforward to calculate an accurate temperature-dependent estimate of $S_2$.

In general, performing $N$ replica simulations enables the unbiased sampling of $\rho^N$ and thus $S_N$. $S_2$ is particularly simple computationally because, as can be seen from Eq.~(\ref{r2_dmqmc}), only amplitudes from psips on the same sites need to be multiplied.  Note that, unlike in Ref.~\onlinecite{Hastings2010}, the replica trick is used here only to ensure the statistical independence of the two reduced density matrices.

\section{Importance sampling}
\label{sec:importance_sampling}
This article considers the application of DMQMC to the $S=1/2$ antiferromagnetic Heisenberg model,
\begin{equation}
\hat{H} = J \: \sum_{\langle i,j \rangle} \boldsymbol{\hat{S}}_i \cdot \boldsymbol{\hat{S}}_j,
\label{heisenberg}
\end{equation}
where $J>0$ and the $\langle i,j \rangle$ implies that the summation is over nearest-neighbor pairs of spins only. Periodic boundary conditions are applied. We work in the standard basis set where the many-spin states are tensor products of the one-spin eigenstates, $\ket{\uparrow}$ and $\ket{\downarrow}$, of the $\hat{S}_z$ operator.

The algorithm described in section II allows a stochastic sampling of the exact finite-temperature density matrix, assuming $\Delta \beta$ is sufficiently small. However, for the antiferromagnetic Heisenberg model, the sampling method described is found to be insufficient for all but the smallest systems.  The ground-state wave function is highly delocalized over the states in the basis set, and thus the ground-state density matrix, $\rho = \ket{E_0}\bra{E_0}$, has many off-diagonal elements with magnitudes comparable to the diagonal elements.  As a result, when the density matrix is sampled via the DMQMC algorithm, only a small fraction of psips reside on or near diagonal elements.  Estimators for the expectation values of most operators of interest only receive contributions from psips on or near the diagonal, so very few psips contribute and these estimators suffer from large statistical errors at low temperatures.  Importance sampling can greatly improve the sampling quality.  

We start by defining the excitation level between basis states $\ket{X_i}$ and $\ket{X_j}$ to be the smallest number of pairs of opposite spins that must be flipped in order to reach $\ket{X_i}$ from $\ket{X_j}$. For the Heisenberg model, Eq.~(\ref{heisenberg}), the excitation level can change by at most $\pm1$ in a single application of the Hamiltonian.

One straightforward way to improve the quality of sampling is to reduce the probability of psips spawning far from the diagonal of the density matrix. Psips
that \emph{do} reside on higher excitation levels are given a correspondingly larger weight, so that expectation values of operators are unchanged. However, the increase in the population of low-weight psips near the diagonal of the density matrix reduces the stochastic error in near-diagonal expectation values.

With this motivation in mind, we define the following importance-sampling procedure; a more rigorous formulation is given in Appendix~\ref{appendix:importancesampling}.
Every time a psip on excitation level $\gamma$ attempts to spawn a new psip on excitation level $\delta$, the probability of successful spawning is altered by a factor $P_{\gamma\delta}$. Thus, if a psip on a diagonal element attempts to spawn a new psip onto the first excitation level, the probability of successful spawning is altered by a factor $P_{01}$, where $P_{01}<1$. 
The first excitation level will thus be occupied by $P_{01}$ times as many psips as it would have been with unaltered spawning probabilities.
The reduced (relative) population must be accounted for by giving all psips in the first excitation level a weight of $W_1=1/P_{01}$ when evaluating estimators.
The number of spawning attempts from the first to the second excitation level is also altered by 
a factor $P_{01}$ (due to the reduced number of psips at level 1) and the chance of successful spawning for each attempt is multiplied by the factor $P_{12}$. Hence, the weight given to the second excitation level must be $W_2=1/P_{01}P_{12}$.
Furthermore, since there are $P_{01}$ times as many psips on the first excitation level, the probability of spawning from the first excitation level to the diagonal elements must be enhanced by a factor $1/P_{01}$ in order to achieve consistent spawning dynamics.
\begin{figure}
\includegraphics{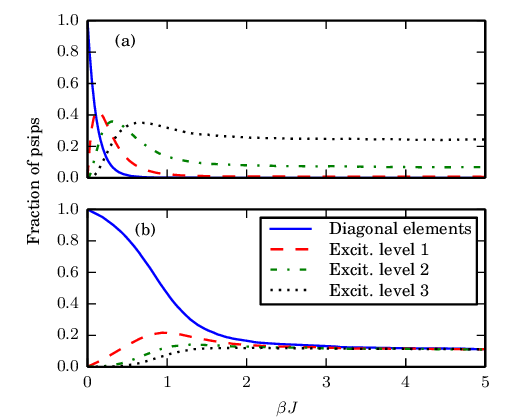}
\caption{(Color online.) The inverse-temperature dependence of the fractions of psips on the first four excitation levels during a DMQMC simulation of a 4$\times$4 square antiferromagnetic Heisenberg lattice\cite{suppl_mat}. A single $\beta$-loop was carried out using an initial population of $10^5$ psips on the diagonal elements.  The results shown in (a) were obtained without importance sampling. The results shown in (b) were obtained using importance sampling to ensure that every excitation level had a roughly equal number of psips at zero temperature.}
\label{fig:importance_sampling}
\end{figure}

In general $P_{\gamma\delta} = 1/P_{\delta\gamma}$ and the weight given to a psip on excitation level $\gamma$ is
\begin{equation}
W_{\gamma} = \prod_{\delta=1}^{\delta=\gamma} \frac{1}{P_{\delta-1,\delta}}.
\label{eq:weights}
\end{equation}
The estimator for an expectation value, $\langle\hat{O}\rangle$, is thus
\begin{equation}
\langle \hat{O} \rangle = \frac{\sum_{i,j} \tilde{\rho}_{ij}W_{E(i,j)} O_{ji}}{\sum_i \tilde{\rho}_{ii}},
\end{equation}
where $E(i,j)$ is the excitation level between $\ket{X_i}$ and $\ket{X_j}$, and $\tilde{\rho}_{ij}$ is the importance-sampled density matrix. The expected value of the importance-sampled psip charge on site $(i,j)$ is proportional to $\tilde{\rho}_{ij}$. 

There is clearly some freedom in the numerical values chosen for the factors $P_{\gamma\delta}$.  In this study we adjust $P_{\gamma\delta}$ such that all excitation levels have similar psip populations in the ground state. Whilst this choice is not necessarily optimal, it successfully increases the quality of ground-state estimates by many orders of magnitude whilst still allowing the entire density matrix to be sampled.
Figures~\ref{fig:importance_sampling}(a) and~\ref{fig:importance_sampling}(b) show the fractions of psips on the first four excitation levels for the $4\times4$ square Heisenberg lattice as obtained without and with importance sampling.
Both simulations used an initial population of $10^5$ psips and a single $\beta$-loop.  It is clear that importance sampling greatly assists in keeping a non-negligible population on the lower excitation levels.

In practice, the values of $P_{\gamma\delta}$ are very small for small $\gamma$ and $\delta$ and decrease in magnitude as the lattice size increases. In order to avoid an unnecessarily large suppression of psip spawning from the diagonal at high temperatures, we introduce the weights $W_{\gamma}$ and probabilities $P_{\gamma\delta}$ gradually from $\beta=0$ until they reach their desired final values at $\beta_{\textrm{target}}$. For $\beta < \beta_{\textrm{target}}$, the weight at excitation level $\gamma$ is set equal to $(W_\gamma)^{\beta/\beta_{\textrm{target}}}$; for $\beta > \beta_{\textrm{target}}$, the weight is held constant. The value of $\beta_{\textrm{target}}$ is chosen to be the inverse temperature at which the ground state is deemed to have been reached. However, the value of this parameter is not critical for the quality of the sampling.

\section{Results}
\label{sec:results}

Results are presented for the $S=1/2$ antiferromagnetic Heisenberg model. In order to calculate finite-temperature properties, it is necessary to include contributions from all $M_S$ (total spin) subspaces. Because the Heisenberg Hamiltonian conserves $M_S$, different subspaces can be studied using separate simulations for estimators of the form in Eq.~(\ref{eq:estimator}), which is an embarrassingly parallel computational task. Combining such results for different values of $M_S$ is straightforward but requires additional calculations and reveals nothing of interest about DMQMC\@. The energy and staggered magnetisation results presented in this paper are for the $M_S=0$ subspace only. For more general quantities such as $S_2$ it is necessary to perform a simulation with psips spanning all $M_S$ subspaces simultaneously.

\subsection{Temperature-dependent energy on the $4 \times 4$ lattice}
\begin{figure}
\includegraphics{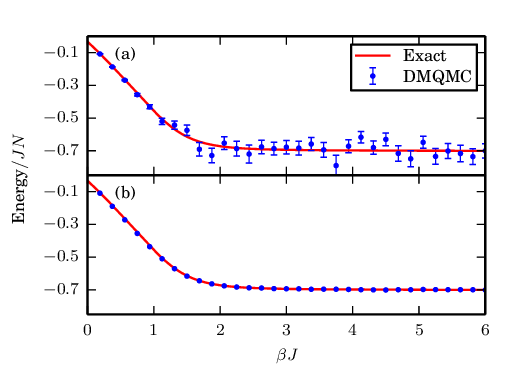}
\caption{(Color online.) The exact and DMQMC energies for the 4$\times$4 antiferromagnetic Heisenberg lattice (Hilbert space dimension $D \approx 1.29 \times 10^{4}$) with periodic boundary conditions~\cite{suppl_mat}.  The $\beta$ step size, $\Delta \beta$, obeys $JN\Delta \beta=0.1$, where $N=16$, and results are shown every 30 iterations. Each simulation consisted of 1000 $\beta$-loops.  For the simulation reported in (a), 100 psips were introduced at the start of each $\beta$-loop; for (b), $10^5$ psips were introduced. The error bars in (b) are smaller than the size of the markers.}
\label{fig:4x4_energy}
\end{figure}
The 4$\times$4 lattice is small enough that an exact diagonalization can be performed, thus allowing a direct check of our DMQMC results.  Figure~\ref{fig:4x4_energy}(a) shows the energy as a function of temperature using an initial population of 100 psips at the start of each $\beta$-loop and accumulating statistics over 1000 such loops.  The shift was allowed to vary throughout the simulation and there were typically 700--800 psips in the simulation by the end of each $\beta$-loop. No importance sampling was applied, and so statistical flucuations increase with inverse temperature as explained in Sec.~\ref{sec:importance_sampling}.  Increasing the initial population to $10^5$ psips (Fig.~\ref{fig:4x4_energy}(b)) reduces the statistical errors such that the agreement with the exact results is essentially perfect.

\subsection{Temperature-dependent staggered magnetisation on the $8 \times 8$ lattice}
\begin{figure}
\includegraphics{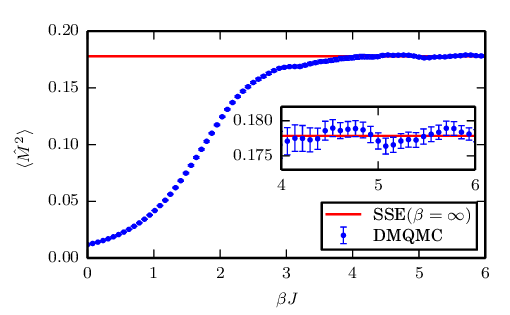}
\caption{(Color online.) The square of the staggered magnetization for an 8$\times$8 antiferromagnetic Heisenberg lattice ($D \approx 1.83 \times 10^{18}$) using $1.4 \times 10^7$ psips and 143 $\beta$-loops~\cite{suppl_mat}. The importance-sampling procedure described in the main text was applied. The DMQMC value of $\braket{\hat{M}^2}$ is plotted every 50th iteration.  Error bars, where not visible, are smaller than the size of the marker.  The ground-state value of $\braket{\hat{M}^2}$ obtained from an SSE simulation~\cite{Sandvik1997} is plotted for comparison.}
\label{fig:8x8_plot}
\end{figure}

The square of the staggered magnetization is represented by the operator
\begin{equation}
\hat{M}^2 = \boldsymbol{\hat{M}} \cdot\boldsymbol{\hat{M}}, \; \; \; \textrm{with} \; \boldsymbol{\hat{M}}  = \frac{1}{N} \sum_i (-1)^{x_i + y_i} \boldsymbol{\hat{S}}_i,
\label{eq:M2}
\end{equation}
where $x_i$ and $x_j$ denote the coordinates of the square lattice.  This operator does not commute with the Hamiltonian. Its expectation value, $\braket{\hat{M}^2}$, is plotted as a function of $\beta$ in Fig.~\ref{fig:8x8_plot} for an 8$\times$8 lattice; a ground-state value obtained using the SSE method~\cite{Sandvik1997} is also shown. The dimension of the $M_S=0$ Hilbert space of an 8$\times$8 Heisenberg Hamiltonian is approximately $1.83 \times 10^{18}$, so the density matrix sampled in this simulation has approximately $3.36 \times 10^{36}$ elements.

\subsection{Ground-state energy and staggered magnetisation on the $6 \times6$ lattice}
For ground-state calculations statistics are accumulated over a single $\beta$-loop after the ground state has been reached.  A relatively modest simulation of a 6$\times$6 lattice using $6.5 \times 10^6$ psips gave a ground-state energy of $-0.67888(24)JN$ and a value of $\braket{\hat{M}^2}$ equal to $0.20985(7)$, where errors were obtained via a blocking analysis~\cite{Flyvbjerg1989}. These values agree with exact results~\cite{m2_note} of $-0.678872JN$ and $0.20983$ for the ground-state energy and staggered magnetization, respectively, to within statistical errors~\cite{Schulz1996}. Remarkably, despite the fact that $[\hat{M}^2,\hat{H}] \neq 0$, the statistical error in the value of $\braket{\hat{M}^2}$ was smaller than that of the energy. It was not necessary to reweight these results using Eq.~(\ref{bias_weights}) as the psip population was large enough to render such biases negligible.

\subsection{Ground-state concurrence on spin rings}
The ground-state concurrence, $\mathcal{C}_{gs}$, for neighboring spins (qubits) on an antiferromagnetic Heisenberg ring was studied by Wootters and O'Connor in 2001~\cite{OConnor2001}. They showed that, for an even number of spins, $\mathcal{C}_{gs}$ has a simple relationship with the ground-state energy,
\begin{equation}
\mathcal{C}_{gs} = -\frac{1}{2}(4E_0/N + 1).
\label{eq:gsc}
\end{equation}
The exact results calculated from this formula provide a useful test of our DMQMC estimates of $\mathcal{C}_{gs}$, which are obtained from the sampled reduced density matrix (see Appendix~\ref{appendix:concurrence}).

The DMQMC estimates of $\mathcal{C}_{gs}$ are presented in Table~\ref{tab:concurrence}, along with Wootters and O'Connor's exact analytic values for up to $N=10$ sites. A ring with $N=36$ sites was also studied, using FCIQMC to calculate $E_0$ and then Eq.~(\ref{eq:gsc}) to obtain a value of $\mathcal{C}_{gs}$ for comparison with the DMQMC results. For lengths up to $N=10$, the calculation of the concurrence can easily be carried out using other methods and provides a straightforward test of the DMQMC algorithm. The $N=36$ chain is far from trivial and it is promising that such accurate results can be obtained.

We note that the DMQMC estimates of $\mathcal{C}_{gs}$ were obtained by sampling the reduced density matrix for a single pair of spins. Due to translational invariance, these estimates could have been improved by sampling the reduced density matrix for every neighboring pair and combining the results.
\begin{table}
\begin{center}
{\footnotesize
\begin{tabular}{ccccccc}
\hline
\hline
& & \multicolumn{2}{c}{$\mathcal{C}_{gs}$} & & &  \\
\cline{3-4}
$N$ & $E_0/N$ & Exact~\cite{OConnor2001} & DMQMC & $N_p$ & $N_l$ & $D$ \\
\hline
4 & -0.5000 & 0.5000 & 0.5005(4) & $2.5 \times 10^3$ & 250 & 6 \\
6 & -0.4671 & 0.4343 & 0.4342(5) & $2.5 \times 10^3$ & $1 \times 10^3$ & 20 \\
8 & -0.4564 & 0.4128 & 0.4129(5) & $1 \times 10^4$ & $1 \times 10^3$ & 70 \\
10 & -0.4515 & 0.4031 & 0.4031(4) & $4 \times 10^3$ & $1 \times 10^3$ & 252  \\
36 & -0.44374(2) & 0.38748(4) & 0.3873(8) & $1 \times 10^6$ & 12 & $9.08\times10^9$ \\
\hline
\hline
\end{tabular}
}
\caption{DMQMC estimates of the ground-state concurrence, $\mathcal{C}_{gs}$, for antiferromagnetic spin rings containing $N$ sites in the absence of an external magnetic field.   Each ring correponds to a Hilbert space of $D$ basis functions.  The density matrix was sampled using $N_p$ psips, and statistics accumulated over $N_l$ $\beta$-loops.  Exact results for the energy and concurrence for rings up to $N=10$ are taken from Ref.~(\onlinecite{OConnor2001}).  An FCIQMC calculation was used to find the energy of the $N=36$ chain, from which an ``exact'' value of $C_{gs}$ was determined using Eq.~(\ref{eq:gsc}). The full density matrix of the $N=36$ chain has approximately $D^2 = 8.24 \times 10^{19}$ elements.}
\label{tab:concurrence}
\end{center}
\end{table}

\subsection{Temperature-dependent Renyi-2 entropy on the $6 \times 6$ lattice}

\begin{figure}
\includegraphics{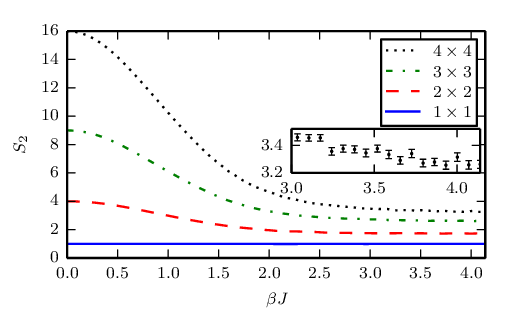}
\caption{(Color online.) Renyi-2 entropy for square sublattices of a $6\times6$ antiferromagnetic Heisenberg lattice. Error bars, where not shown, are smaller than the line thickness.  The inset zooms in on the $4\times4$ sublattice, which has the largest error bars. $S_2$ values were taken from every 25th iteration and averaged over $16\text{--}180$ loops over the temperature range.}
\label{fig:square_sublattices}
\end{figure}

\begin{figure}
\includegraphics{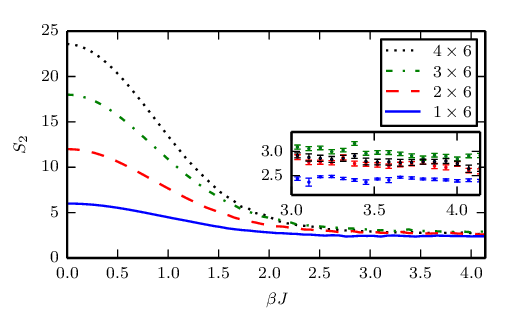}
\caption{(Color online.) Renyi-2 entropy for strip sublattices of a $6\times6$ lattice.  Error bars are shown in the inset (color scheme as in the main plot) for large $\beta$, where they are largest, and are smaller than the linewidth where not visible. $S_2$ values were taken from every 25th iteration and averaged over $32\text{--}240$ loops over the temperature range.  As expected, $S_2$ for the $2\times6$ and $4\times6$ sublattices tend to the same value in the zero-temperature limit.}
\label{fig:strip_sublattices}
\end{figure}

Temperature-dependent Renyi-2 entropies were calculated for various sublattices of the $6\times6$ lattice using the replica approach described in Section~\ref{sec:DMQMC}. Translational invariance was used to improve the quality of the statistics and importance sampling was once again applied. Note that, unlike other calculations presented, the initial population spanned the entire Hilbert space rather than just the $M_S=0$ subspace in order to capture the correct behaviour of $S_2$ at non-zero temperatures.
Results are presented in Figure~\ref{fig:square_sublattices} for square sublattices and in Figure~\ref{fig:strip_sublattices} for strip sublattices. We were unable to find results in the literature for comparison.

\section{The sign problem in DMQMC}
\label{sec:sign_problem}

As discussed in Sec.~\ref{sec:FCIQMC}, the annihilation step in FCIQMC leads to the characteristic population dynamics, whereby the sign problem can only be overcome once a critical psip population (the `plateau') has been exceeded.
Since the annihilation steps in FCIQMC and DMQMC are identical, we would expect DMQMC to possess similar population dynamics.
The annihilation rate for a given psip population in DMQMC will be significantly smaller than in FCIQMC because the number of density matrix elements is the square of the number of basis functions. As such, it is to be expected that the plateau height for a given Hamiltonian will be higher in DMQMC than in FCIQMC.

\begin{figure}
\includegraphics{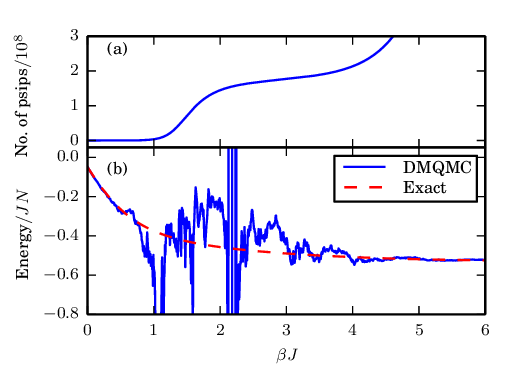}
\caption{(Color online.) Population dynamics and energy estimate for the antiferromagnetic Heisenberg model on a 4$\times$4 triangular lattice with periodic boundary conditions. A fixed shift was used and a single $\beta$-loop performed. Panel (a) shows the emergence of a plateau in the psip population and (b) shows the exact energy and the DMQMC estimate of the energy as functions of $\beta$. Once the plateau has been exited, the DMQMC energy is in good agreement with the exact results.  The severity of the sign problem increases with inverse temperature.}
\label{fig:plateau}
\end{figure}

To investigate this issue, we considered the antiferromagnetic Heisenberg model on a 4$\times$4 triangular lattice, for which an exact diagonalization is easily performed. The triangular lattice is the archetypal example of a frustrated lattice and has a severe sign problem.  We carried out a single $\beta$-loop and allowed the population to grow with a fixed shift whilst simultaneously investigating the accuracy of the DMQMC energy estimate. Figure~\ref{fig:plateau}(a) demonstrates that the population plateaus as expected. Figure~\ref{fig:plateau}(b) shows the accuracy of the energy estimate throughout the plateau period. 

At high temperatures accurate results are obtained. This is also the case in other finite-temperature methods, where it is generally found that the sign problem is less severe for small $\beta$; indeed, there is no sign problem at all at infinite temperature.  However, at lower temperatures the energy estimates suffer large fluctuations, and it is not until the population exceeds the plateau that a good agreement with exact results is once again obtained.  The plateau height occurs at $\sim 1.75\times10^8$ psips. For comparison, the plateau population in an FCIQMC calculation of the same system is at $\sim 2.0\times10^4$ psips. We find (in this case) that the DMQMC plateau height is approximately the square of the plateau height in FCIQMC, matching the increase in the size of the space being sampled.  It is possible to obtain accurate results for the entire temperature range simply by starting the simulation with an initial population greater than that of the plateau. Moreover, even if the plateau cannot be reached due to memory restrictions, one can nevertheless systematically reach lower temperatures by increasing the population of psips.

The sign problem in the antiferromagnetic Heisenberg model on a triangular lattice is severe in both FCIQMC and DMQMC~\cite{Spencer2012}.  However, the efficiency of the annihilation procedure in FCIQMC varies substantially with the system studied and with the basis set used~\cite{Spencer2012}. Furthermore, the initiator approximation (i-FCIQMC) proposed by Cleland \emph{et al.}~\cite{Cleland2010} has been shown to reduce the memory requirements for FCIQMC calculations by several orders of magnitude in many cases. In i-FCIQMC, spawning events onto previously unoccupied sites are forbidden unless the psip population of the parent site exceeds a threshold. This increases the annihilation rate relative to FCIQMC and ameliorates the sign problem at the expense of introducing a systematically improvable approximation. We have yet to investigate the DMQMC equivalent of the initiator method. It is expected that ground-state properties will be available, but it is not yet clear to what extent the modified spawning will affect finite-temperature results.

\section{Discussion}
\label{sec:discussion}

This article has described DMQMC, a quantum Monte Carlo method that allows direct sampling of the finite-temperature and ground-state density matrices in a discrete basis. The validity of the method has been verified by reproducing exact and well-established results for some small systems, including the calculation of the concurrence of one-dimensional spin rings by directly sampling reduced density matrices. In all cases investigated, DMQMC has proved capable and accurate.

The introduction of an importance-sampling procedure allows larger lattices to be investigated --- the largest system that we have simulated successfully to date is a 10$\times$10 antiferromagnetic Heisenberg model. Larger systems could easily be tackled at high temperatures, but it is unlikely that our very simple approach to importance sampling will allow simulations of lattices of more than 10$\times$10 sites over the entire range of temperatures from infinity to zero. If a better importance-sampling procedure can be devised, as we believe likely, many more avenues will be opened.

Like FCIQMC, DMQMC uses an annihilation procedure that allows it to overcome the sign problem if a system-specific population of psips is reached. In FCIQMC the sign problem can often be ameliorated, for example by changing to a more appropriate basis set~\cite{Spencer2012,Kolodrubetz2013} or by applying the initiator approximation~\cite{Cleland2010}.  Given the similarities between DMQMC and FCIQMC, it is likely that these ideas and future developments in FCIQMC will also apply to DMQMC\@. Due to the relative youth of FCIQMC, the rate of theoretical and algorithmic improvements is rapid~\cite{Cleland2010, Thom2010, Petruzielo2012, Mukherjee2013, Roggero2013}.

Obtaining expectation values of operators that are non-linear in the full density matrix or functions of a reduced density matrix is challenging.
We used the replica approach~\cite{Hastings2010}, in which two DMQMC simulations are performed side-by-side, to obtain unbiased estimates of the Renyi-2 entropy for several subsystems of a Heisenberg lattice across a range of temperatures.

The replica approach might also be useful in FCIQMC for unbiased sampling of expectation values of operators that do not commute with the Hamiltonian.
However, a general calculation of this type would require running over every pair of psips in both calculations, which would be computationally demanding and have a major impact on the efficient parallelisation of the FCIQMC algorithm.

The lattices studied with DMQMC so far are small by the standards of
path-integral quantum Monte Carlo methods. However, the unique and defining
feature of DMQMC is that it samples the full density matrix. This allows the calculation of arbitrary expectation values and quantum information measures inaccessible using other quantum Monte Carlo methods. We wish to finish by emphasising that, whilst the applications presented here have been to lattice models, the formulation of DMQMC is general.
For example, DMQMC could be used to calculate the two-electron reduced density matrix for small molecules, which could then be used to construct accurate exchange-correlation functionals for density functional theory~\cite{Hood97}.
 
\begin{acknowledgments}
    We are grateful for enlightening discussions with Prof.~Terry Rudolph, Dr.~Alex Thom and Tom Poole.
    This work made use of the Imperial College London High Performance Computing facilities.
    J.S.S. and W.M.C.F. acknowledge the stimulating research environment provided by the Thomas Young Centre under grant number TYC-101.
    N.S.B. was supported by an Undergraduate Research Opportunities Scholarship in the Centre for Doctoral Training on Theory and Simulation of Materials at Imperial College funded by EPSRC under grant number EP/G036888/1.
\end{acknowledgments}

\appendix

\section{Concurrence}
\label{appendix:concurrence}

For the special case of a two-qubit state, the entanglement of formation can be calculated from a quantity known as the ``concurrence''. Given a reduced density matrix $\bm{\rho}_A$, where in this case $A$ refers to a subsystem of two qubits, the concurrence is defined as
\begin{equation}
\mathcal{C}(\bm{\rho}_A) \equiv \textnormal{max}(0, \gamma_1-\gamma_2-\gamma_3-\gamma_4),
\end{equation}
where $\gamma_1 > \gamma_2 > \gamma_3 > \gamma_4$ are the eigenvalues of the matrix
\begin{equation}
\bm{R} = \sqrt{\sqrt{\bm{\rho}_A}\tilde{\bm{\rho}}_A\sqrt{\bm{\rho}_A}}
\end{equation}
and 
\begin{equation}
\tilde{\bm{\rho}}_A = (\sigma_y\otimes\sigma_y)\bm{\rho}_A^*(\sigma_y\otimes\sigma_y),
\end{equation}
with $\sigma_y$ the Pauli spin matrix for the $y$-direction. This expression is only valid in the standard basis set.

The value of the concurrence $\mathcal{C}$ ranges from zero to one and is monotonically related to the entanglement of formation~\cite{Hill1997}; the concurrence can therefore be regarded as a measure of entanglement. The entanglement of formation for two qubits~\cite{Wootters2001} is given by
\begin{equation}
\mathcal{E}(\mathcal{C}) = h\left( \frac{1+\sqrt{1-\mathcal{C}^2}}{2}\right),
\end{equation}
where
\begin{equation}
h(x) = -x\log_2 x - (1-x)\log_2 (1-x).
\end{equation}

If the Hamiltonian and thus the reduced density matrix are real (as is the case in the Heisenberg model), the calculation of $\mathcal{C}$ reduces to the calculation of the moduli of the eigenvalues of 
\begin{equation}
\label{eq:concurrenceMatrix}
\bm{R} = \bm{\rho}_A(\sigma_y\otimes\sigma_y).
\end{equation}
Since $\bm{R}$ is a $4\times4$ matrix, it is trivial to compute the concurrence by direct diagonalization once $\bm{\rho}_A$ is known.

Using DMQMC, it is possible to sample the unnormalized reduced density matrix as described in Sec.~\ref{sec:DMQMC} and thus to estimate the concurrence using
\begin{equation}
\langle \mathcal{C} \rangle = \frac{\textnormal{max}(0,\gamma_1-\gamma_2-\gamma_3-\gamma_4)}{\textnormal{Tr}(\boldsymbol{\rho}_A)},
\end{equation}
where $\boldsymbol{\rho}_A$ is now the unnormalized reduced density matrix and $\{\gamma_i\}$ are the eigenvalues of the unnormalized matrix $\bm{R}$.

\section{Importance sampling}
\label{appendix:importancesampling}

The DMQMC evolution equation is
\begin{equation}
\frac{d \rho_{ij}}{d\beta} = \frac{1}{2} \sum_{k} (T_{ik}\rho_{kj} + \rho_{ik} T_{kj}),
\label{eq:evolution_equation_before}
\end{equation}
where $\rho_{ij}$ is an element of $\hat{\rho}$ in the many-particle basis chosen for the simulation. Instead of $\rho_{ij}$, we would like to sample the importance-sampled density matrix
\begin{equation}
\tilde{\rho}_{ij} = \frac{\rho_{ij}}{W_{E(i,j)}},
\end{equation}
where $E(i,j)$ is the excitation level of the pair $(i,j)$ and $W_{\alpha}$ is defined in Eq.~(\ref{eq:weights}). Following the standard procedure of importance sampling, we introduce a trial function
\begin{equation}
\rho_{ij}^T = \frac{1}{W_{E(i,j)}}.
\end{equation}
This matrix is symmetric, $\rho_{ij}^T = \rho_{ji}^T$, as $E(i,j)=E(j,i)$. The importance-sampled density matrix then has components $\tilde{\rho}_{ij} = \rho_{ij}^T \rho_{ij}$, where \emph{no} summation is performed over indices. Multiplying Eq.~(\ref{eq:evolution_equation_before}) by $\rho_{ij}^T$ yields 
the following evolution equation for $\tilde{\rho}_{ij}$:
\begin{align}
\frac{d (\rho_{ij}^T \rho_{ij})}{d\beta} &= \frac{1}{2} \sum_{k} (\rho_{ij}^T T_{ik}\rho_{kj} + \rho_{ik} \rho_{ij}^T T_{kj}) \\
\begin{split}
                                         &= \frac{1}{2} \sum_{k} \left[ \left(\rho_{ij}^T T_{ik}\frac{1}{\rho_{kj}^T}\right) \left(\rho_{kj}^T \rho_{kj} \right) \right. \\
                                         & \phantom{000000000} \left. + \left(\rho_{ik}^T  \rho_{ik}\right) \left(\rho_{ij}^T T_{kj}\frac{1}{\rho_{ik}^T}\right) \right] \\
\end{split}
\end{align}                                         
\begin{equation}                                         
\Rightarrow \frac{d\tilde{\rho}_{ij}}{d\beta} = \frac{1}{2} \sum_{k} \left[ (\rho_{ij}^T T_{ik}\frac{1}{\rho_{kj}^T}) \tilde{\rho}_{kj} + \tilde{\rho}_{ik} (\rho_{ij}^T T_{kj}\frac{1}{\rho_{ik}^T}) \right]. \label{eq:importance_sampled_evolution}
\end{equation}
The above differential equation is entirely analogous to Eq.~(\ref{eq:symmetrisedBlochWithUpdateMatix}). As such, the finite-difference version of Eq.~(\ref{eq:importance_sampled_evolution}) can be simulated in an almost identical manner to the standard DMQMC algorithm: the extra factors of $\rho^T$ simply act to alter the spawning probabilities.  Consider the case where the excitation level of $(i,j)$ is $\gamma$ and excitation level of $(k,j)$ is $\gamma-1$. The probability that a spawning attempt from $(k,j)$ to $(i,j)$ is successful is altered by the factor
\begin{equation}
    \rho_{ij}^T \frac{1}{\rho_{kj}^T} = \frac{W_{E(k,j)}}{W_{E(i,j)}}
    = \frac{W_{\gamma-1}}{W_{\gamma}} = P_{\gamma-1,\gamma},
\end{equation}
where Eq.~(\ref{eq:weights}) has been used to simplify the expression. Similarly, when spawning in the opposite direction, the probability is altered by the reciprocal of this factor. Spawning events that do not alter the excitation level are unaffected.

\begin{thebibliography}{53}%
\makeatletter
\providecommand \@ifxundefined [1]{%
 \@ifx{#1\undefined}
}%
\providecommand \@ifnum [1]{%
 \ifnum #1\expandafter \@firstoftwo
 \else \expandafter \@secondoftwo
 \fi
}%
\providecommand \@ifx [1]{%
 \ifx #1\expandafter \@firstoftwo
 \else \expandafter \@secondoftwo
 \fi
}%
\providecommand \natexlab [1]{#1}%
\providecommand \enquote  [1]{``#1''}%
\providecommand \bibnamefont  [1]{#1}%
\providecommand \bibfnamefont [1]{#1}%
\providecommand \citenamefont [1]{#1}%
\providecommand \href@noop [0]{\@secondoftwo}%
\providecommand \href [0]{\begingroup \@sanitize@url \@href}%
\providecommand \@href[1]{\@@startlink{#1}\@@href}%
\providecommand \@@href[1]{\endgroup#1\@@endlink}%
\providecommand \@sanitize@url [0]{\catcode `\\12\catcode `\$12\catcode
  `\&12\catcode `\#12\catcode `\^12\catcode `\_12\catcode `\%12\relax}%
\providecommand \@@startlink[1]{}%
\providecommand \@@endlink[0]{}%
\providecommand \url  [0]{\begingroup\@sanitize@url \@url }%
\providecommand \@url [1]{\endgroup\@href {#1}{\urlprefix }}%
\providecommand \urlprefix  [0]{URL }%
\providecommand \Eprint [0]{\href }%
\providecommand \doibase [0]{http://dx.doi.org/}%
\providecommand \selectlanguage [0]{\@gobble}%
\providecommand \bibinfo  [0]{\@secondoftwo}%
\providecommand \bibfield  [0]{\@secondoftwo}%
\providecommand \translation [1]{[#1]}%
\providecommand \BibitemOpen [0]{}%
\providecommand \bibitemStop [0]{}%
\providecommand \bibitemNoStop [0]{.\EOS\space}%
\providecommand \EOS [0]{\spacefactor3000\relax}%
\providecommand \BibitemShut  [1]{\csname bibitem#1\endcsname}%
\let\auto@bib@innerbib\@empty
\bibitem [{\citenamefont {Hammond}\ \emph {et~al.}(1994)\citenamefont
  {Hammond}, \citenamefont {Lester~Jr.},\ and\ \citenamefont
  {J.}}]{Hammond1994}%
  \BibitemOpen
  \bibfield  {author} {\bibinfo {author} {\bibfnamefont {B.~L.}\ \bibnamefont
  {Hammond}}, \bibinfo {author} {\bibfnamefont {W.~A.}\ \bibnamefont
  {Lester~Jr.}}, \ and\ \bibinfo {author} {\bibfnamefont {R.~P.}\ \bibnamefont
  {J.}},\ }\href@noop {} {\emph {\bibinfo {title} {Monte Carlo Methods in Ab
  Initio Quantum Chemistry}}}\ (\bibinfo  {publisher} {World Scientific},\
  \bibinfo {address} {Singapore},\ \bibinfo {year} {1994})\BibitemShut
  {NoStop}%
\bibitem [{\citenamefont {Foulkes}\ \emph {et~al.}(2001)\citenamefont
  {Foulkes}, \citenamefont {Mitas}, \citenamefont {Needs},\ and\ \citenamefont
  {Rajagopal}}]{Foulkes2001}%
  \BibitemOpen
  \bibfield  {author} {\bibinfo {author} {\bibfnamefont {W.~M.~C.}\
  \bibnamefont {Foulkes}}, \bibinfo {author} {\bibfnamefont {L.}~\bibnamefont
  {Mitas}}, \bibinfo {author} {\bibfnamefont {R.~J.}\ \bibnamefont {Needs}}, \
  and\ \bibinfo {author} {\bibfnamefont {G.}~\bibnamefont {Rajagopal}},\
  }\href@noop {} {\bibfield  {journal} {\bibinfo  {journal} {Rev. Mod. Phys.}\
  }\textbf {\bibinfo {volume} {73}},\ \bibinfo {pages} {33} (\bibinfo {year}
  {2001})}\BibitemShut {NoStop}%
\bibitem [{\citenamefont {Kalos}(1962)}]{Kalos1962}%
  \BibitemOpen
  \bibfield  {author} {\bibinfo {author} {\bibfnamefont {M.~H.}\ \bibnamefont
  {Kalos}},\ }\href@noop {} {\bibfield  {journal} {\bibinfo  {journal} {Phys.
  Rev.}\ }\textbf {\bibinfo {volume} {128}},\ \bibinfo {pages} {1791} (\bibinfo
  {year} {1962})}\BibitemShut {NoStop}%
\bibitem [{\citenamefont {Ceperley}\ and\ \citenamefont
  {Kalos}(1979)}]{Ceperley1979}%
  \BibitemOpen
  \bibfield  {author} {\bibinfo {author} {\bibfnamefont {D.~M.}\ \bibnamefont
  {Ceperley}}\ and\ \bibinfo {author} {\bibfnamefont {M.~H.}\ \bibnamefont
  {Kalos}},\ }in\ \href@noop {} {\emph {\bibinfo {booktitle} {{Monte Carlo
  Methods in Statistical Physics}}}},\ \bibinfo {editor} {edited by\ \bibinfo
  {editor} {\bibfnamefont {K.}~\bibnamefont {Binder}}}\ (\bibinfo  {publisher}
  {Springer-Verlag},\ \bibinfo {year} {1979})\ pp.\ \bibinfo {pages}
  {145--194}\BibitemShut {NoStop}%
\bibitem [{\citenamefont {Anderson}(1975)}]{Anderson1975}%
  \BibitemOpen
  \bibfield  {author} {\bibinfo {author} {\bibfnamefont {J.~B.}\ \bibnamefont
  {Anderson}},\ }\href@noop {} {\bibfield  {journal} {\bibinfo  {journal} {J.
  Chem. Phys.}\ }\textbf {\bibinfo {volume} {63}},\ \bibinfo {pages} {1499}
  (\bibinfo {year} {1975})}\BibitemShut {NoStop}%
\bibitem [{\citenamefont {Anderson}(1976)}]{Anderson1976}%
  \BibitemOpen
  \bibfield  {author} {\bibinfo {author} {\bibfnamefont {J.~B.}\ \bibnamefont
  {Anderson}},\ }\href@noop {} {\bibfield  {journal} {\bibinfo  {journal} {J.
  Chem. Phys.}\ }\textbf {\bibinfo {volume} {65}},\ \bibinfo {pages} {4121}
  (\bibinfo {year} {1976})}\BibitemShut {NoStop}%
\bibitem [{\citenamefont {Moskowitz}\ \emph {et~al.}(1982)\citenamefont
  {Moskowitz}, \citenamefont {Schmidt}, \citenamefont {Lee},\ and\
  \citenamefont {Kalos}}]{Moskowitz1982}%
  \BibitemOpen
  \bibfield  {author} {\bibinfo {author} {\bibfnamefont {J.~W.}\ \bibnamefont
  {Moskowitz}}, \bibinfo {author} {\bibfnamefont {K.~E.}\ \bibnamefont
  {Schmidt}}, \bibinfo {author} {\bibfnamefont {M.~A.}\ \bibnamefont {Lee}}, \
  and\ \bibinfo {author} {\bibfnamefont {M.~H.}\ \bibnamefont {Kalos}},\
  }\href@noop {} {\bibfield  {journal} {\bibinfo  {journal} {J. Chem. Phys.}\
  }\textbf {\bibinfo {volume} {77}},\ \bibinfo {pages} {349} (\bibinfo {year}
  {1982})}\BibitemShut {NoStop}%
\bibitem [{\citenamefont {Reynolds}\ \emph {et~al.}(1982)\citenamefont
  {Reynolds}, \citenamefont {Ceperley}, \citenamefont {Alder},\ and\
  \citenamefont {Lester~Jr.}}]{Reynolds1982}%
  \BibitemOpen
  \bibfield  {author} {\bibinfo {author} {\bibfnamefont {P.~J.}\ \bibnamefont
  {Reynolds}}, \bibinfo {author} {\bibfnamefont {D.~M.}\ \bibnamefont
  {Ceperley}}, \bibinfo {author} {\bibfnamefont {B.~J.}\ \bibnamefont {Alder}},
  \ and\ \bibinfo {author} {\bibfnamefont {W.~A.}\ \bibnamefont {Lester~Jr.}},\
  }\href@noop {} {\bibfield  {journal} {\bibinfo  {journal} {J. Chem. Phys.}\
  }\textbf {\bibinfo {volume} {77}},\ \bibinfo {pages} {5593} (\bibinfo {year}
  {1982})}\BibitemShut {NoStop}%
\bibitem [{\citenamefont {Gaudoin}\ and\ \citenamefont
  {Pitarke}(2007)}]{Gaudoin2007}%
  \BibitemOpen
  \bibfield  {author} {\bibinfo {author} {\bibfnamefont {R.}~\bibnamefont
  {Gaudoin}}\ and\ \bibinfo {author} {\bibfnamefont {J.~M.}\ \bibnamefont
  {Pitarke}},\ }\href@noop {} {\bibfield  {journal} {\bibinfo  {journal} {Phys.
  Rev. Lett.}\ }\textbf {\bibinfo {volume} {99}},\ \bibinfo {pages} {126406}
  (\bibinfo {year} {2007})}\BibitemShut {NoStop}%
\bibitem [{\citenamefont {Casulleras}\ and\ \citenamefont
  {Boronat}(1995)}]{Casulleras95}%
  \BibitemOpen
  \bibfield  {author} {\bibinfo {author} {\bibfnamefont {J.}~\bibnamefont
  {Casulleras}}\ and\ \bibinfo {author} {\bibfnamefont {J.}~\bibnamefont
  {Boronat}},\ }\href {\doibase 10.1103/PhysRevB.52.3654} {\bibfield  {journal}
  {\bibinfo  {journal} {Phys. Rev. B}\ }\textbf {\bibinfo {volume} {52}},\
  \bibinfo {pages} {3654} (\bibinfo {year} {1995})}\BibitemShut {NoStop}%
\bibitem [{\citenamefont {Feynman}(1953)}]{Feynman1953}%
  \BibitemOpen
  \bibfield  {author} {\bibinfo {author} {\bibfnamefont {R.~P.}\ \bibnamefont
  {Feynman}},\ }\href@noop {} {\bibfield  {journal} {\bibinfo  {journal} {Phys.
  Rev.}\ }\textbf {\bibinfo {volume} {91}},\ \bibinfo {pages} {1291} (\bibinfo
  {year} {1953})}\BibitemShut {NoStop}%
\bibitem [{\citenamefont {Sandvik}(2010)}]{Sandvik2010}%
  \BibitemOpen
  \bibfield  {author} {\bibinfo {author} {\bibfnamefont {A.~W.}\ \bibnamefont
  {Sandvik}},\ }\href@noop {} {\bibfield  {journal} {\bibinfo  {journal} {AIP
  Conference Proceedings}\ }\textbf {\bibinfo {volume} {1297}},\ \bibinfo
  {pages} {135} (\bibinfo {year} {2010})}\BibitemShut {NoStop}%
\bibitem [{\citenamefont {Booth}\ \emph {et~al.}(2009)\citenamefont {Booth},
  \citenamefont {Thom},\ and\ \citenamefont {Alavi}}]{Booth2009}%
  \BibitemOpen
  \bibfield  {author} {\bibinfo {author} {\bibfnamefont {G.~H.}\ \bibnamefont
  {Booth}}, \bibinfo {author} {\bibfnamefont {A.~J.~W.}\ \bibnamefont {Thom}},
  \ and\ \bibinfo {author} {\bibfnamefont {A.}~\bibnamefont {Alavi}},\ }\href
  {\doibase 10.1063/1.3193710} {\bibfield  {journal} {\bibinfo  {journal} {J.
  Chem. Phys.}\ }\textbf {\bibinfo {volume} {131}},\ \bibinfo {pages} {054106}
  (\bibinfo {year} {2009})}\BibitemShut {NoStop}%
\bibitem [{\citenamefont {Booth}\ \emph {et~al.}(2012)\citenamefont {Booth},
  \citenamefont {Gruneis}, \citenamefont {Kresse},\ and\ \citenamefont
  {Alavi}}]{Booth2013}%
  \BibitemOpen
  \bibfield  {author} {\bibinfo {author} {\bibfnamefont {G.~H.}\ \bibnamefont
  {Booth}}, \bibinfo {author} {\bibfnamefont {A.}~\bibnamefont {Gruneis}},
  \bibinfo {author} {\bibfnamefont {G.}~\bibnamefont {Kresse}}, \ and\ \bibinfo
  {author} {\bibfnamefont {A.}~\bibnamefont {Alavi}},\ }\href {\doibase
  10.1038/nature11770} {\bibfield  {journal} {\bibinfo  {journal} {Nature}\
  }\textbf {\bibinfo {volume} {493}},\ \bibinfo {pages} {365} (\bibinfo {year}
  {2012})}\BibitemShut {NoStop}%
\bibitem [{\citenamefont {Spencer}\ \emph {et~al.}(2012)\citenamefont
  {Spencer}, \citenamefont {Blunt},\ and\ \citenamefont
  {Foulkes}}]{Spencer2012}%
  \BibitemOpen
  \bibfield  {author} {\bibinfo {author} {\bibfnamefont {J.~S.}\ \bibnamefont
  {Spencer}}, \bibinfo {author} {\bibfnamefont {N.~S.}\ \bibnamefont {Blunt}},
  \ and\ \bibinfo {author} {\bibfnamefont {W.~M.~C.}\ \bibnamefont {Foulkes}},\
  }\href@noop {} {\bibfield  {journal} {\bibinfo  {journal} {J. Chem. Phys.}\
  }\textbf {\bibinfo {volume} {136}},\ \bibinfo {pages} {054110} (\bibinfo
  {year} {2012})}\BibitemShut {NoStop}%
\bibitem [{\citenamefont {Booth}\ and\ \citenamefont
  {Alavi}(2010)}]{Booth2010}%
  \BibitemOpen
  \bibfield  {author} {\bibinfo {author} {\bibfnamefont {G.~H.}\ \bibnamefont
  {Booth}}\ and\ \bibinfo {author} {\bibfnamefont {A.}~\bibnamefont {Alavi}},\
  }\href {\doibase 10.1063/1.3407895} {\bibfield  {journal} {\bibinfo
  {journal} {J. Chem. Phys.}\ }\textbf {\bibinfo {volume} {132}},\ \bibinfo
  {pages} {174104} (\bibinfo {year} {2010})}\BibitemShut {NoStop}%
\bibitem [{\citenamefont {Cleland}\ \emph {et~al.}(2010)\citenamefont
  {Cleland}, \citenamefont {Booth},\ and\ \citenamefont {Alavi}}]{Cleland2010}%
  \BibitemOpen
  \bibfield  {author} {\bibinfo {author} {\bibfnamefont {D.~M.}\ \bibnamefont
  {Cleland}}, \bibinfo {author} {\bibfnamefont {G.~H.}\ \bibnamefont {Booth}},
  \ and\ \bibinfo {author} {\bibfnamefont {A.}~\bibnamefont {Alavi}},\ }\href
  {\doibase 10.1063/1.3302277} {\bibfield  {journal} {\bibinfo  {journal} {J.
  Chem. Phys.}\ }\textbf {\bibinfo {volume} {132}},\ \bibinfo {pages} {041103}
  (\bibinfo {year} {2010})}\BibitemShut {NoStop}%
\bibitem [{\citenamefont {Cleland}\ \emph {et~al.}(2012)\citenamefont
  {Cleland}, \citenamefont {Booth}, \citenamefont {Overy},\ and\ \citenamefont
  {Alavi}}]{Cleland2012}%
  \BibitemOpen
  \bibfield  {author} {\bibinfo {author} {\bibfnamefont {D.}~\bibnamefont
  {Cleland}}, \bibinfo {author} {\bibfnamefont {G.~H.}\ \bibnamefont {Booth}},
  \bibinfo {author} {\bibfnamefont {C.}~\bibnamefont {Overy}}, \ and\ \bibinfo
  {author} {\bibfnamefont {A.}~\bibnamefont {Alavi}},\ }\href {\doibase
  10.1021/ct300504f} {\bibfield  {journal} {\bibinfo  {journal} {J. Chem.
  Theor. Comput.}\ }\textbf {\bibinfo {volume} {8}},\ \bibinfo {pages} {4138}
  (\bibinfo {year} {2012})}\BibitemShut {NoStop}%
\bibitem [{\citenamefont {Shepherd}\ \emph {et~al.}(2012)\citenamefont
  {Shepherd}, \citenamefont {Booth}, \citenamefont {Gr\"uneis},\ and\
  \citenamefont {Alavi}}]{Shepherd2012PRB}%
  \BibitemOpen
  \bibfield  {author} {\bibinfo {author} {\bibfnamefont {J.~J.}\ \bibnamefont
  {Shepherd}}, \bibinfo {author} {\bibfnamefont {G.}~\bibnamefont {Booth}},
  \bibinfo {author} {\bibfnamefont {A.}~\bibnamefont {Gr\"uneis}}, \ and\
  \bibinfo {author} {\bibfnamefont {A.}~\bibnamefont {Alavi}},\ }\href
  {\doibase 10.1103/PhysRevB.85.081103} {\bibfield  {journal} {\bibinfo
  {journal} {Phys. Rev. B}\ }\textbf {\bibinfo {volume} {85}},\ \bibinfo
  {pages} {081103} (\bibinfo {year} {2012})}\BibitemShut {NoStop}%
\bibitem [{\citenamefont {Daday}\ \emph {et~al.}(2012)\citenamefont {Daday},
  \citenamefont {Smart}, \citenamefont {Booth}, \citenamefont {Alavi},\ and\
  \citenamefont {Filippi}}]{Daday2012}%
  \BibitemOpen
  \bibfield  {author} {\bibinfo {author} {\bibfnamefont {C.}~\bibnamefont
  {Daday}}, \bibinfo {author} {\bibfnamefont {S.}~\bibnamefont {Smart}},
  \bibinfo {author} {\bibfnamefont {G.~H.}\ \bibnamefont {Booth}}, \bibinfo
  {author} {\bibfnamefont {A.}~\bibnamefont {Alavi}}, \ and\ \bibinfo {author}
  {\bibfnamefont {C.}~\bibnamefont {Filippi}},\ }\href {\doibase
  10.1021/ct300486d} {\bibfield  {journal} {\bibinfo  {journal} {J. Chem.
  Theor. Comput.}\ }\textbf {\bibinfo {volume} {8}},\ \bibinfo {pages} {4441}
  (\bibinfo {year} {2012})}\BibitemShut {NoStop}%
\bibitem [{\citenamefont {Petruzielo}\ \emph {et~al.}(2012)\citenamefont
  {Petruzielo}, \citenamefont {Holmes}, \citenamefont {Changlani},
  \citenamefont {Nightingale},\ and\ \citenamefont {Umrigar}}]{Petruzielo2012}%
  \BibitemOpen
  \bibfield  {author} {\bibinfo {author} {\bibfnamefont {F.~R.}\ \bibnamefont
  {Petruzielo}}, \bibinfo {author} {\bibfnamefont {A.~A.}\ \bibnamefont
  {Holmes}}, \bibinfo {author} {\bibfnamefont {H.~J.}\ \bibnamefont
  {Changlani}}, \bibinfo {author} {\bibfnamefont {M.~P.}\ \bibnamefont
  {Nightingale}}, \ and\ \bibinfo {author} {\bibfnamefont {C.~J.}\ \bibnamefont
  {Umrigar}},\ }\href {\doibase 10.1103/PhysRevLett.109.230201} {\bibfield
  {journal} {\bibinfo  {journal} {Phys. Rev. Lett.}\ }\textbf {\bibinfo
  {volume} {109}},\ \bibinfo {pages} {230201} (\bibinfo {year}
  {2012})}\BibitemShut {NoStop}%
\bibitem [{\citenamefont {Kalos}\ and\ \citenamefont
  {Pederiva}(2000)}]{PhysRevLett.85.3547}%
  \BibitemOpen
  \bibfield  {author} {\bibinfo {author} {\bibfnamefont {M.~H.}\ \bibnamefont
  {Kalos}}\ and\ \bibinfo {author} {\bibfnamefont {F.}~\bibnamefont
  {Pederiva}},\ }\href {\doibase 10.1103/PhysRevLett.85.3547} {\bibfield
  {journal} {\bibinfo  {journal} {Phys. Rev. Lett.}\ }\textbf {\bibinfo
  {volume} {85}},\ \bibinfo {pages} {3547} (\bibinfo {year}
  {2000})}\BibitemShut {NoStop}%
\bibitem [{\citenamefont {Carlson}\ and\ \citenamefont
  {Kalos}(1985)}]{PhysRevC.32.1735}%
  \BibitemOpen
  \bibfield  {author} {\bibinfo {author} {\bibfnamefont {J.}~\bibnamefont
  {Carlson}}\ and\ \bibinfo {author} {\bibfnamefont {M.~H.}\ \bibnamefont
  {Kalos}},\ }\href {\doibase 10.1103/PhysRevC.32.1735} {\bibfield  {journal}
  {\bibinfo  {journal} {Phys. Rev. C}\ }\textbf {\bibinfo {volume} {32}},\
  \bibinfo {pages} {1735} (\bibinfo {year} {1985})}\BibitemShut {NoStop}%
\bibitem [{\citenamefont {Anderson}\ \emph {et~al.}(1991)\citenamefont
  {Anderson}, \citenamefont {Traynor},\ and\ \citenamefont
  {Boghosian}}]{anderson:7418}%
  \BibitemOpen
  \bibfield  {author} {\bibinfo {author} {\bibfnamefont {J.~B.}\ \bibnamefont
  {Anderson}}, \bibinfo {author} {\bibfnamefont {C.~A.}\ \bibnamefont
  {Traynor}}, \ and\ \bibinfo {author} {\bibfnamefont {B.~M.}\ \bibnamefont
  {Boghosian}},\ }\href {\doibase 10.1063/1.461368} {\bibfield  {journal}
  {\bibinfo  {journal} {J. Chem. Phys.}\ }\textbf {\bibinfo {volume} {95}},\
  \bibinfo {pages} {7418} (\bibinfo {year} {1991})}\BibitemShut {NoStop}%
\bibitem [{\citenamefont {Zhang}\ and\ \citenamefont
  {Kalos}(1991)}]{PhysRevLett.67.3074}%
  \BibitemOpen
  \bibfield  {author} {\bibinfo {author} {\bibfnamefont {S.}~\bibnamefont
  {Zhang}}\ and\ \bibinfo {author} {\bibfnamefont {M.~H.}\ \bibnamefont
  {Kalos}},\ }\href {\doibase 10.1103/PhysRevLett.67.3074} {\bibfield
  {journal} {\bibinfo  {journal} {Phys. Rev. Lett.}\ }\textbf {\bibinfo
  {volume} {67}},\ \bibinfo {pages} {3074} (\bibinfo {year}
  {1991})}\BibitemShut {NoStop}%
\bibitem [{\citenamefont {Liu}\ \emph {et~al.}(1994)\citenamefont {Liu},
  \citenamefont {Zhang},\ and\ \citenamefont {Kalos}}]{PhysRevE.50.3220}%
  \BibitemOpen
  \bibfield  {author} {\bibinfo {author} {\bibfnamefont {Z.}~\bibnamefont
  {Liu}}, \bibinfo {author} {\bibfnamefont {S.}~\bibnamefont {Zhang}}, \ and\
  \bibinfo {author} {\bibfnamefont {M.~H.}\ \bibnamefont {Kalos}},\ }\href
  {\doibase 10.1103/PhysRevE.50.3220} {\bibfield  {journal} {\bibinfo
  {journal} {Phys. Rev. E}\ }\textbf {\bibinfo {volume} {50}},\ \bibinfo
  {pages} {3220} (\bibinfo {year} {1994})}\BibitemShut {NoStop}%
\bibitem [{\citenamefont {Kalos}(1996)}]{PhysRevE.53.5420}%
  \BibitemOpen
  \bibfield  {author} {\bibinfo {author} {\bibfnamefont {M.~H.}\ \bibnamefont
  {Kalos}},\ }\href {\doibase 10.1103/PhysRevE.53.5420} {\bibfield  {journal}
  {\bibinfo  {journal} {Phys. Rev. E}\ }\textbf {\bibinfo {volume} {53}},\
  \bibinfo {pages} {5420} (\bibinfo {year} {1996})}\BibitemShut {NoStop}%
\bibitem [{Note1()}]{Note1}%
  \BibitemOpen
  \bibinfo {note} {There is a close relationship between the symmetrized Bloch
  equation in Eq.~(\ref {eq:symmetrisedBloch}) and the von Neumann and quantum
  Liouville equations, $i\hbar \protect \frac {\partial \rho }{\partial t} =
  [H,\rho ]$. The von Neumann equation may be derived by using the
  time-dependent Schr{\" o}dinger equation to work out how the arbitrary
  initial density matrix of Eq.~(\ref {eq:rho_mixed}) evolves. Analogously, the
  Bloch equation may be derived by using the \protect \emph {imaginary-time}
  Schr{\" o}dinger equation to work out how Eq.~(\ref {eq:rho_mixed}) evolves
  in imaginary time. The switch from real to imaginary time replaces the
  commutator of $\protect \mathaccentV {hat}05E{H}$ and $\rho $ appearing in
  the von Neumann equation with the anticommutator appearing in the Bloch
  equation. The additional factor of $\protect \frac {1}{2}$ is introduced for
  convenience: it ensures that evolving Eq.~(\ref {eq:symmetrisedBloch}) for
  $\beta $ units of imaginary time, starting from the unit operator, $\rho
  (\beta =0)=\protect \mathaccentV {hat}05E{I}$, yields the canonical density
  matrix at inverse temperature $\beta $, not $2\beta $.}\BibitemShut {Stop}%
\bibitem [{\citenamefont {Umrigar}\ \emph {et~al.}(1993)\citenamefont
  {Umrigar}, \citenamefont {Nightingale},\ and\ \citenamefont
  {Runge}}]{Umrigar1993}%
  \BibitemOpen
  \bibfield  {author} {\bibinfo {author} {\bibfnamefont {C.~J.}\ \bibnamefont
  {Umrigar}}, \bibinfo {author} {\bibfnamefont {M.~P.}\ \bibnamefont
  {Nightingale}}, \ and\ \bibinfo {author} {\bibfnamefont {K.~J.}\ \bibnamefont
  {Runge}},\ }\href {\doibase 10.1063/1.465195} {\bibfield  {journal} {\bibinfo
   {journal} {J. Chem. Phys.}\ }\textbf {\bibinfo {volume} {99}},\ \bibinfo
  {pages} {2865} (\bibinfo {year} {1993})}\BibitemShut {NoStop}%
\bibitem [{\citenamefont {Amico}\ \emph {et~al.}(2008)\citenamefont {Amico},
  \citenamefont {Fazio}, \citenamefont {Osterloh},\ and\ \citenamefont
  {Vedral}}]{Amico2008}%
  \BibitemOpen
  \bibfield  {author} {\bibinfo {author} {\bibfnamefont {L.}~\bibnamefont
  {Amico}}, \bibinfo {author} {\bibfnamefont {R.}~\bibnamefont {Fazio}},
  \bibinfo {author} {\bibfnamefont {A.}~\bibnamefont {Osterloh}}, \ and\
  \bibinfo {author} {\bibfnamefont {V.}~\bibnamefont {Vedral}},\ }\href@noop {}
  {\bibfield  {journal} {\bibinfo  {journal} {Rev. Mod. Phys.}\ }\textbf
  {\bibinfo {volume} {80}},\ \bibinfo {pages} {517 } (\bibinfo {year}
  {2008})}\BibitemShut {NoStop}%
\bibitem [{\citenamefont {McMinis}\ and\ \citenamefont
  {Tubman}(2013)}]{McMinis2013}%
  \BibitemOpen
  \bibfield  {author} {\bibinfo {author} {\bibfnamefont {J.}~\bibnamefont
  {McMinis}}\ and\ \bibinfo {author} {\bibfnamefont {N.~M.}\ \bibnamefont
  {Tubman}},\ }\href {\doibase 10.1103/PhysRevB.87.081108} {\bibfield
  {journal} {\bibinfo  {journal} {Phys. Rev. B}\ }\textbf {\bibinfo {volume}
  {87}},\ \bibinfo {pages} {081108} (\bibinfo {year} {2013})}\BibitemShut
  {NoStop}%
\bibitem [{\citenamefont {Swingle}\ \emph {et~al.}(2013)\citenamefont
  {Swingle}, \citenamefont {McMinis},\ and\ \citenamefont
  {Tubman}}]{Swingle2013}%
  \BibitemOpen
  \bibfield  {author} {\bibinfo {author} {\bibfnamefont {B.}~\bibnamefont
  {Swingle}}, \bibinfo {author} {\bibfnamefont {J.}~\bibnamefont {McMinis}}, \
  and\ \bibinfo {author} {\bibfnamefont {N.~M.}\ \bibnamefont {Tubman}},\
  }\href {\doibase 10.1103/PhysRevB.87.235112} {\bibfield  {journal} {\bibinfo
  {journal} {Phys. Rev. B}\ }\textbf {\bibinfo {volume} {87}},\ \bibinfo
  {pages} {235112} (\bibinfo {year} {2013})}\BibitemShut {NoStop}%
\bibitem [{\citenamefont {Tubman}\ and\ \citenamefont
  {McMinis}(2012)}]{Tubman2012}%
  \BibitemOpen
  \bibfield  {author} {\bibinfo {author} {\bibfnamefont {N.~M.}\ \bibnamefont
  {Tubman}}\ and\ \bibinfo {author} {\bibfnamefont {J.}~\bibnamefont
  {McMinis}},\ }\href@noop {} {\  (\bibinfo {year} {2012})},\ \Eprint
  {http://arxiv.org/abs/1204.4731} {arXiv:1204.4731 [cond-mat.str-el]}
  \BibitemShut {NoStop}%
\bibitem [{\citenamefont {Hastings}\ \emph {et~al.}(2010)\citenamefont
  {Hastings}, \citenamefont {Gonz\'{a}lez}, \citenamefont {Kallin},\ and\
  \citenamefont {Melko}}]{Hastings2010}%
  \BibitemOpen
  \bibfield  {author} {\bibinfo {author} {\bibfnamefont {M.~B.}\ \bibnamefont
  {Hastings}}, \bibinfo {author} {\bibfnamefont {I.}~\bibnamefont
  {Gonz\'{a}lez}}, \bibinfo {author} {\bibfnamefont {A.~B.}\ \bibnamefont
  {Kallin}}, \ and\ \bibinfo {author} {\bibfnamefont {R.~G.}\ \bibnamefont
  {Melko}},\ }\href@noop {} {\bibfield  {journal} {\bibinfo  {journal} {Phys.
  Rev. Lett.}\ }\textbf {\bibinfo {volume} {104}},\ \bibinfo {pages} {157201}
  (\bibinfo {year} {2010})}\BibitemShut {NoStop}%
\bibitem [{\citenamefont {Glaser}\ \emph {et~al.}(2003)\citenamefont {Glaser},
  \citenamefont {B\"{u}ttner},\ and\ \citenamefont {Fehske}}]{Glaser2003}%
  \BibitemOpen
  \bibfield  {author} {\bibinfo {author} {\bibfnamefont {U.}~\bibnamefont
  {Glaser}}, \bibinfo {author} {\bibfnamefont {H.}~\bibnamefont {B\"{u}ttner}},
  \ and\ \bibinfo {author} {\bibfnamefont {H.}~\bibnamefont {Fehske}},\
  }\href@noop {} {\bibfield  {journal} {\bibinfo  {journal} {Phys. Rev. A}\
  }\textbf {\bibinfo {volume} {68}},\ \bibinfo {pages} {032318} (\bibinfo
  {year} {2003})}\BibitemShut {NoStop}%
\bibitem [{\citenamefont {Roscilde}\ \emph {et~al.}(2005)\citenamefont
  {Roscilde}, \citenamefont {Verrucchi}, \citenamefont {Fubini}, \citenamefont
  {Haas},\ and\ \citenamefont {Tognetti}}]{Roscilde2005}%
  \BibitemOpen
  \bibfield  {author} {\bibinfo {author} {\bibfnamefont {T.}~\bibnamefont
  {Roscilde}}, \bibinfo {author} {\bibfnamefont {P.}~\bibnamefont {Verrucchi}},
  \bibinfo {author} {\bibfnamefont {A.}~\bibnamefont {Fubini}}, \bibinfo
  {author} {\bibfnamefont {S.}~\bibnamefont {Haas}}, \ and\ \bibinfo {author}
  {\bibfnamefont {V.}~\bibnamefont {Tognetti}},\ }\href@noop {} {\bibfield
  {journal} {\bibinfo  {journal} {Phys. Rev. Lett.}\ }\textbf {\bibinfo
  {volume} {94}},\ \bibinfo {pages} {147208} (\bibinfo {year}
  {2005})}\BibitemShut {NoStop}%
\bibitem [{\citenamefont {Sandvik}(2005)}]{Sandvik2005}%
  \BibitemOpen
  \bibfield  {author} {\bibinfo {author} {\bibfnamefont {A.~W.}\ \bibnamefont
  {Sandvik}},\ }\href@noop {} {\bibfield  {journal} {\bibinfo  {journal} {Phys.
  Rev. Lett}\ }\textbf {\bibinfo {volume} {95}},\ \bibinfo {pages} {207203}
  (\bibinfo {year} {2005})}\BibitemShut {NoStop}%
\bibitem [{\citenamefont {Alet}\ \emph {et~al.}(2007)\citenamefont {Alet},
  \citenamefont {Capponi}, \citenamefont {Laflorencie},\ and\ \citenamefont
  {Mambrini}}]{Alet2007}%
  \BibitemOpen
  \bibfield  {author} {\bibinfo {author} {\bibfnamefont {F.}~\bibnamefont
  {Alet}}, \bibinfo {author} {\bibfnamefont {S.}~\bibnamefont {Capponi}},
  \bibinfo {author} {\bibfnamefont {N.}~\bibnamefont {Laflorencie}}, \ and\
  \bibinfo {author} {\bibfnamefont {M.}~\bibnamefont {Mambrini}},\ }\href@noop
  {} {\bibfield  {journal} {\bibinfo  {journal} {Phys. Rev. Lett.}\ }\textbf
  {\bibinfo {volume} {99}},\ \bibinfo {pages} {117204} (\bibinfo {year}
  {2007})}\BibitemShut {NoStop}%
\bibitem [{\citenamefont {Lin}\ and\ \citenamefont {Sandvik}(2010)}]{Lin2010}%
  \BibitemOpen
  \bibfield  {author} {\bibinfo {author} {\bibfnamefont {Y.-C.}\ \bibnamefont
  {Lin}}\ and\ \bibinfo {author} {\bibfnamefont {A.~W.}\ \bibnamefont
  {Sandvik}},\ }\href@noop {} {\bibfield  {journal} {\bibinfo  {journal} {Phys.
  Rev. B}\ }\textbf {\bibinfo {volume} {82}},\ \bibinfo {pages} {224414}
  (\bibinfo {year} {2010})}\BibitemShut {NoStop}%
\bibitem [{\citenamefont {Knowles}\ and\ \citenamefont
  {Handy}(1984)}]{Knowles1984}%
  \BibitemOpen
  \bibfield  {author} {\bibinfo {author} {\bibfnamefont {P.}~\bibnamefont
  {Knowles}}\ and\ \bibinfo {author} {\bibfnamefont {N.}~\bibnamefont
  {Handy}},\ }\href {\doibase 10.1016/0009-2614(84)85513-X} {\bibfield
  {journal} {\bibinfo  {journal} {Chem. Phys. Lett.}\ }\textbf {\bibinfo
  {volume} {111}},\ \bibinfo {pages} {315 } (\bibinfo {year}
  {1984})}\BibitemShut {NoStop}%
\bibitem [{sup()}]{suppl_mat}%
  \BibitemOpen
  \href@noop {} {}\bibinfo {note} {See Supplemental Material at URL (to be
  inserted by publisher) and doi:10.6084/m9.figshare.655964 for raw
  data.}\BibitemShut {Stop}%
\bibitem [{\citenamefont {Sandvik}(1997)}]{Sandvik1997}%
  \BibitemOpen
  \bibfield  {author} {\bibinfo {author} {\bibfnamefont {A.~W.}\ \bibnamefont
  {Sandvik}},\ }\href@noop {} {\bibfield  {journal} {\bibinfo  {journal} {Phys.
  Rev. B}\ }\textbf {\bibinfo {volume} {56}},\ \bibinfo {pages} {11678}
  (\bibinfo {year} {1997})}\BibitemShut {NoStop}%
\bibitem [{\citenamefont {Flybjerg}\ and\ \citenamefont
  {Petersen}(1989)}]{Flyvbjerg1989}%
  \BibitemOpen
  \bibfield  {author} {\bibinfo {author} {\bibfnamefont {H.}~\bibnamefont
  {Flybjerg}}\ and\ \bibinfo {author} {\bibfnamefont {H.}~\bibnamefont
  {Petersen}},\ }\href@noop {} {\bibfield  {journal} {\bibinfo  {journal} {J.
  Chem. Phys.}\ }\textbf {\bibinfo {volume} {91}},\ \bibinfo {pages} {461}
  (\bibinfo {year} {1989})}\BibitemShut {NoStop}%
\bibitem [{m2_()}]{m2_note}%
  \BibitemOpen
  \href@noop {} {}\bibinfo {note} {The definitions of $M^2$ given in
  Eq.~(\ref{eq:M2}) and in Ref.~\onlinecite{Schulz1996} have slightly different
  normalization factors. The result quoted from Ref.~\onlinecite{Schulz1996}
  has been renormalised accordingly.}\BibitemShut {Stop}%
\bibitem [{\citenamefont {Schulz}\ \emph {et~al.}(1996)\citenamefont {Schulz},
  \citenamefont {Ziman},\ and\ \citenamefont {Poilblanc}}]{Schulz1996}%
  \BibitemOpen
  \bibfield  {author} {\bibinfo {author} {\bibfnamefont {H.~J.}\ \bibnamefont
  {Schulz}}, \bibinfo {author} {\bibfnamefont {T.~A.~L.}\ \bibnamefont
  {Ziman}}, \ and\ \bibinfo {author} {\bibfnamefont {D.}~\bibnamefont
  {Poilblanc}},\ }\href@noop {} {\bibfield  {journal} {\bibinfo  {journal}
  {Journal de Physique}\ }\textbf {\bibinfo {volume} {6}},\ \bibinfo {pages}
  {675} (\bibinfo {year} {1996})}\BibitemShut {NoStop}%
\bibitem [{\citenamefont {O'Connor}\ and\ \citenamefont
  {Wootters}(2001)}]{OConnor2001}%
  \BibitemOpen
  \bibfield  {author} {\bibinfo {author} {\bibfnamefont {K.~M.}\ \bibnamefont
  {O'Connor}}\ and\ \bibinfo {author} {\bibfnamefont {W.~K.}\ \bibnamefont
  {Wootters}},\ }\href {\doibase 10.1103/PhysRevA.63.052302} {\bibfield
  {journal} {\bibinfo  {journal} {Phys. Rev. A}\ }\textbf {\bibinfo {volume}
  {63}},\ \bibinfo {pages} {052302} (\bibinfo {year} {2001})}\BibitemShut
  {NoStop}%
\bibitem [{\citenamefont {Kolodrubetz}\ \emph {et~al.}(2013)\citenamefont
  {Kolodrubetz}, \citenamefont {Spencer}, \citenamefont {Clark},\ and\
  \citenamefont {Foulkes}}]{Kolodrubetz2013}%
  \BibitemOpen
  \bibfield  {author} {\bibinfo {author} {\bibfnamefont {M.~H.}\ \bibnamefont
  {Kolodrubetz}}, \bibinfo {author} {\bibfnamefont {J.~S.}\ \bibnamefont
  {Spencer}}, \bibinfo {author} {\bibfnamefont {B.~K.}\ \bibnamefont {Clark}},
  \ and\ \bibinfo {author} {\bibfnamefont {W.~M.}\ \bibnamefont {Foulkes}},\
  }\href {\doibase 10.1063/1.4773819} {\bibfield  {journal} {\bibinfo
  {journal} {J. Chem. Phys.}\ }\textbf {\bibinfo {volume} {138}},\ \bibinfo
  {eid} {024110} (\bibinfo {year} {2013})}\BibitemShut {NoStop}%
\bibitem [{\citenamefont {Thom}(2010)}]{Thom2010}%
  \BibitemOpen
  \bibfield  {author} {\bibinfo {author} {\bibfnamefont {A.~J.~W.}\
  \bibnamefont {Thom}},\ }\href {\doibase 10.1103/PhysRevLett.105.263004}
  {\bibfield  {journal} {\bibinfo  {journal} {Phys. Rev. Lett.}\ }\textbf
  {\bibinfo {volume} {105}},\ \bibinfo {pages} {263004} (\bibinfo {year}
  {2010})}\BibitemShut {NoStop}%
\bibitem [{\citenamefont {Mukherjee}\ and\ \citenamefont
  {Alhassid}(2013)}]{Mukherjee2013}%
  \BibitemOpen
  \bibfield  {author} {\bibinfo {author} {\bibfnamefont {A.}~\bibnamefont
  {Mukherjee}}\ and\ \bibinfo {author} {\bibfnamefont {Y.}~\bibnamefont
  {Alhassid}},\ }\href {\doibase 10.1103/PhysRevA.88.053622} {\bibfield
  {journal} {\bibinfo  {journal} {Phys. Rev. A}\ }\textbf {\bibinfo {volume}
  {88}},\ \bibinfo {pages} {053622} (\bibinfo {year} {2013})}\BibitemShut
  {NoStop}%
\bibitem [{\citenamefont {Roggero}\ \emph {et~al.}(2013)\citenamefont
  {Roggero}, \citenamefont {Mukherjee},\ and\ \citenamefont
  {Pederiva}}]{Roggero2013}%
  \BibitemOpen
  \bibfield  {author} {\bibinfo {author} {\bibfnamefont {A.}~\bibnamefont
  {Roggero}}, \bibinfo {author} {\bibfnamefont {A.}~\bibnamefont {Mukherjee}},
  \ and\ \bibinfo {author} {\bibfnamefont {F.}~\bibnamefont {Pederiva}},\
  }\href {\doibase 10.1103/PhysRevB.88.115138} {\bibfield  {journal} {\bibinfo
  {journal} {Phys. Rev. B}\ }\textbf {\bibinfo {volume} {88}},\ \bibinfo
  {pages} {115138} (\bibinfo {year} {2013})}\BibitemShut {NoStop}%
\bibitem [{\citenamefont {Hood}\ \emph {et~al.}(1997)\citenamefont {Hood},
  \citenamefont {Chou}, \citenamefont {Williamson}, \citenamefont {Rajagopal},
  \citenamefont {Needs},\ and\ \citenamefont {Foulkes}}]{Hood97}%
  \BibitemOpen
  \bibfield  {author} {\bibinfo {author} {\bibfnamefont {R.~Q.}\ \bibnamefont
  {Hood}}, \bibinfo {author} {\bibfnamefont {M.~Y.}\ \bibnamefont {Chou}},
  \bibinfo {author} {\bibfnamefont {A.~J.}\ \bibnamefont {Williamson}},
  \bibinfo {author} {\bibfnamefont {G.}~\bibnamefont {Rajagopal}}, \bibinfo
  {author} {\bibfnamefont {R.~J.}\ \bibnamefont {Needs}}, \ and\ \bibinfo
  {author} {\bibfnamefont {W.~M.~C.}\ \bibnamefont {Foulkes}},\ }\href
  {\doibase 10.1103/PhysRevLett.78.3350} {\bibfield  {journal} {\bibinfo
  {journal} {Phys. Rev. Lett.}\ }\textbf {\bibinfo {volume} {78}},\ \bibinfo
  {pages} {3350} (\bibinfo {year} {1997})}\BibitemShut {NoStop}%
\bibitem [{\citenamefont {Hill}\ and\ \citenamefont
  {Wootters}(1997)}]{Hill1997}%
  \BibitemOpen
  \bibfield  {author} {\bibinfo {author} {\bibfnamefont {S.}~\bibnamefont
  {Hill}}\ and\ \bibinfo {author} {\bibfnamefont {W.~K.}\ \bibnamefont
  {Wootters}},\ }\href {\doibase 10.1103/PhysRevLett.78.5022} {\bibfield
  {journal} {\bibinfo  {journal} {Phys. Rev. Lett.}\ }\textbf {\bibinfo
  {volume} {78}},\ \bibinfo {pages} {5022} (\bibinfo {year}
  {1997})}\BibitemShut {NoStop}%
\bibitem [{\citenamefont {Wootters}(2001)}]{Wootters2001}%
  \BibitemOpen
  \bibfield  {author} {\bibinfo {author} {\bibfnamefont {W.~K.}\ \bibnamefont
  {Wootters}},\ }\href {http://www.rintonpress.com/journals/qic-1-1/eof2.pdf}
  {\bibfield  {journal} {\bibinfo  {journal} {Quant. Inf. Comp.}\ }\textbf
  {\bibinfo {volume} {1}},\ \bibinfo {pages} {27} (\bibinfo {year}
  {2001})}\BibitemShut {NoStop}%
\end{thebibliography}
%

\end{document}